\documentclass[pra,preprint,preprintnumbers,amsmath,amssymb,showpacs]{revtex4-1}
\usepackage{graphicx}
\begin{document}

\title{Plasmon-assisted electron-electron collisions at metallic surfaces}
\author{Konstantin A. Kouzakov}%
\address{Department of Nuclear Physics and Quantum Theory of Collisions, Faculty of Physics, Lomonosov Moscow State University, Moscow 119991, Russia\\
and Skobeltsyn Institute of Nuclear Physics, Lomonosov Moscow State University, Moscow 119991, Russia}%
\email{kouzakov@srd.sinp.msu.ru}%
\author{Jamal Berakdar}%
\address{Institute of Physics, Martin Luther University Halle-Wittenberg, Halle 06120, Germany}%
\email{Jamal.Berakdar@physik.uni-halle.de}%


\begin{abstract}
We present a theoretical treatment for the ejection of a secondary
electron from a clean metallic surface induced by the impact of a
fast primary electron. Assuming a direct scattering between the
incident, primary electron and the electron in a metal, we
calculate the electron-pair energy distributions at the surfaces
of Al and Be. Different models for the  screening of the
electron-electron interaction are examined and the footprints of
the surface and the bulk plasmon modes are determined and
analyzed. The formulated theoretical approach is compared with the
available experimental data on the electron-pair emission from Al.
\end{abstract}
\maketitle

%
\section{Introduction}
\label{intro}
Much of our knowledge on  the electronic properties  of materials
has been accumulated over the years by studying the spectra of
electrons inelastically reflected from the surfaces of solid
samples. The corresponding technique is usually referred to as
(reflection) electron energy loss spectroscopy
(EELS)~\cite{eels_book}. By bombarding the solid sample by a
monochromatic beam of electrons and measuring their energy loss
and their deflection angle in the reflection mode,
 detailed
information on the collective excitations  of
 surfaces can be extracted, such as  the excitation energies, the lifetimes, and
dispersions.
The energy loss of the impinging electron upon traversing the sample
results in a variety of excitations including
 phonons and plasmon excitations,
interband and intraband transitions, and inner-shell ionizations
(the latter is particularly useful for detecting the elemental
composition of a material).

In the  past two decades qualitative advances have been achieved
in the capabilities of the electron-pair spectroscopy, also called
the $(e,2e)$ spectroscopy, in which one studies the energy and the
angular distributions of two electrons emitted simultaneously from
a surface following the impact by one
electron~\cite{samarin95,schumann05,schumann10}. In essence, the
$(e,2e)$ spectroscopy investigates a specific process that, among
others, also contributes to the EELS signal: the ejection of a
secondary electron induced by the interaction of an impinging,
primary electron with the surface. The basic information that one
can obtain in this way is as follows: (i) the surface one-electron
spin-resolved spectral function
\cite{weigold_book,prl99,prl2000,Halle_PRB2002,schumann10},
 (ii) the mechanisms of electron-electron collisions
at surfaces
\cite{das97,diffPRL,gollisch2001,berakdarSOLSTCOM,serg2006,serg2011},
and (iii) the surface dielectric function
\cite{werner08,werner11,kouzakov03}. Concerning the spectroscopic
data on the surface one-electron states, the $(e,2e)$ method is
close to angle-resolved photoemission spectroscopy
(ARPES)~\cite{arpes,pes} because both measure the energies and the
wave vectors of the surface electrons. An important difference
between the two techniques is that they involve different surface
one-electron transitions. This is due to the fact that in the
$(e,2e)$ case the Coulombic force imposed by the projectile on the
surface electrons is parallel to the momentum
transfer~\cite{fano63} while in the ARPES case the imposed
electric force is, by definition, perpendicular to the momentum
transfer, that is, to the photon momentum. Note in this context
that, in contrast to ARPES, in the $(e,2e)$ method one can vary
the momentum transfer in a wide range. Another marked feature is
the high surface sensitivity of the $(e,2e)$ method, especially in
the grazing-incidence mode, which makes it very promising for
exploring the Shockley~\cite{schockley33} and Tamm~\cite{tamm32}
electronic states found at the surfaces of various materials.

The $(e,2e)$ measurements on LiF films deposited on a Si(001)
surface~\cite{samarin04} gave the first evidence that the $(e,2e)$
spectroscopy is capable of providing insight into the
secondary-electron ejection assisted by a collective excitation in
the surface electronic system. A recent experimental
study~\cite{werner08} evidences a notable enhancement of the
electron-induced secondary-electron emission from an Al(100)
surface at the electron-energy-loss values that are equal to the
bulk and the surface plasmon excitation energies. The above
findings point out the potential for studying directly the
plasmon-assisted electron-electron collisions and, in particular,
the mechanisms of the plasmon decay at surfaces with the $(e,2e)$
method. Moreover, due to its remarkable surface sensitivity the
$(e,2e)$ spectroscopy can be very successful in illuminating the
properties both of conventional surface plasmons supported by
metallic surfaces~\cite{ritchie57} and of the low-energy acoustic
two-dimensional plasmons that have been recently
predicted~\cite{silkin04,pitarke04} and observed~\cite{nature07}.
This calls for the relevant theoretical framework that
consistently incorporates the effects of the surface dielectric
response into the treatment of the $(e,2e)$ process, a task that
hitherto remained outstanding and will be treated in this work.

Two main mechanisms of the secondary-electron ejection from
metallic surfaces following  the electron impact are
possible~\cite{chung77}: (i) due to the direct scattering between
the incident (primary) and the valence-band (or the
conduction-band) electrons and (ii) due to the decay of the bulk
and the surface plasmons excited by the incident electrons. The
bulk-plasmon decay into a single electron-hole pair is governed by
the interband transitions, which in the case of long-wavelength
plasmons are practically vertical in the the reduced-zone scheme.
In a jellium model the interband transitions are absent and, for
instance, within the random phase approximation
(RPA)~\cite{pines53} the decay of a plasmon into a single
electron-hole pair can take place only when its momentum exceeds
some critical value, namely, when the plasmon-dispersion curve
merges with the electron-hole continuum. The simplest scenario for
the direct electron-ejection mechanism assumes a single
electron-electron interaction via the Coulomb potential screened
by the surrounding medium. Dynamical screening effects can result
in a resonant enhancement of the potential at the energy transfers
corresponding to the excitation of collective modes, such as bulk
and surface plasmons. This feature can markedly manifest itself as
a large increase in the yield of the secondary electrons when the
energy loss of a charged projectile is resonant with the plasmon
energy (see, for instance, Ref.~\cite{kouzakov03}). Since the
contributions from both mechanisms exhibit a resonant behavior at
the plasma energies, their separation in the $(e,2e)$ experiment
is not straightforward. On the theoretical side, simple estimates
show that the ratio of the $(e,2e)$ rate due to a direct
scattering to that due to the plasmon decay behaves as
$\propto\Gamma_{pl}^{-1}$, where $\Gamma_{pl}$ is the plasmon line
width. Therefore, the sharper the plasmon resonance is, the more
dominant the role of the direct scattering over that of the
plasmon decay is.

In the present work, we consider theoretically the
plasmon-assisted $(e,2e)$ collisions at the surfaces of the
independent-electron metals and focus on the contribution from the
direct-scattering mechanism. One of the key ingredients that
determine the $(e,2e)$ rate in this case is the dielectric
response of the metallic sample. The presence of the surface
brings about a major complication as the response function
undergoes a sudden change at the metal-vacuum interface. We
address the problem of the surface dielectric response by
considering two well-known approaches. The first one is the
so-called specular-reflection model (SRM), which was first
introduced in Ref.~\cite{srm66} to study surface plasmons. The
other is based on RPA with infinite surface barrier (RPA-IB) for
electrons in metal~\cite{newns70}. Below we incorporate both
approaches in numerical calculations for $(e,2e)$ from Al and Be.
The choice of Al and Be is motivated by the fact that among the
free-electron metals they exhibit respectively a sharp and a wide
plasmon resonance (in terms of the ratio between the plasmon line
width and the plasma energy in these materials). It allows us to
inspect and numerically illustrate the role of plasmon resonances
on the dynamical screening of the electron-electron interaction.

This paper is organized as follows. In Sec.~\ref{gen_form}, we
present basic formulas and approximations for the considered
process. In Sec.~\ref{surf_resp}, models of surface dielectric
response are discussed. Then, in Sec.~\ref{res}, numerical
calculations for Al and Be are presented and analyzed.
Sec.~\ref{th&exp} is devoted to a comparison of the present
theoretical formulation with the recent experimental measurements
on Al~\cite{werner08,werner11}. Finally, conclusions are drawn in
Sec.~\ref{concl}. Unless otherwise stated, atomic units (a.u.,
$e=\hbar=m_e=1$) are used throughout.

\section{General formulation and basic approximations}
\label{gen_form}
\begin{figure}[t]
\begin{center}
\includegraphics[scale=1]{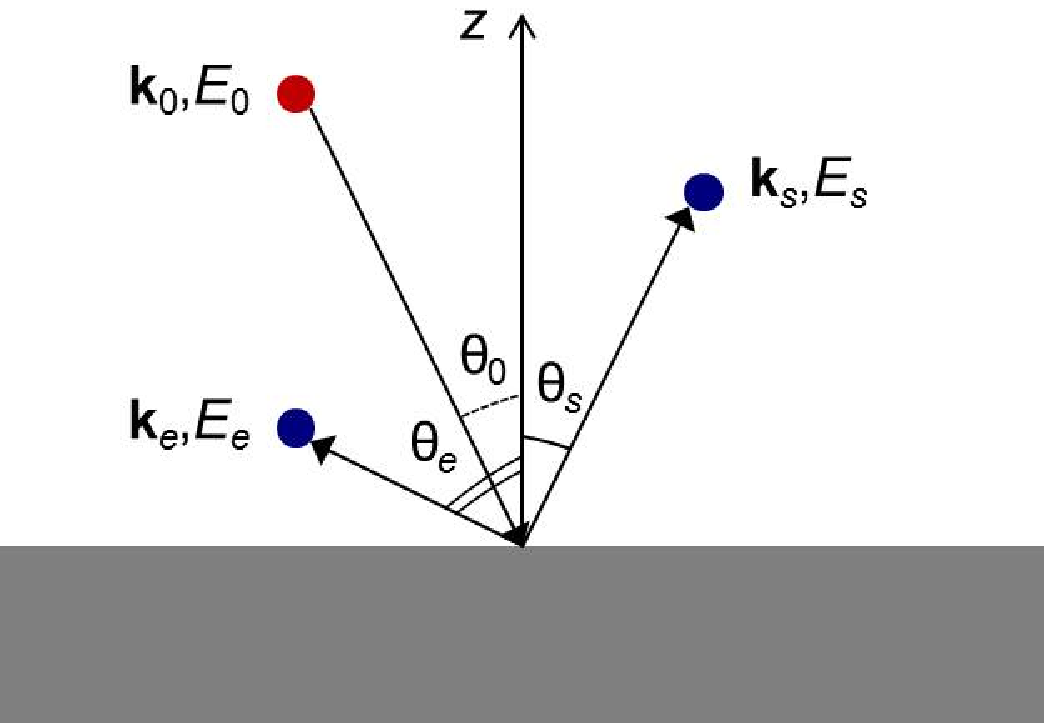}
\end{center}
\caption{\label{setup}(Color online) Schematic drawing of the
electron-induced electron-pair emission from surfaces.}
\end{figure}

We consider the process where, following the impact of a fast
impinging  electron with a momentum ${\bf k}_0$ and energy $E_0$, two
electrons are emitted from the surface of a semiinfinite solid
with momenta ${\bf k}_s$, ${\bf k}_e$ and energies $E_s$, $E_e$
(see Fig.~\ref{setup}). Hereafter the subscript $s~(e)$ stands for
the scattered (ejected) electron. All the energies are measured
with respect to the vacuum level, so that in vacuum we have the dispersion
$$
k_j=\sqrt{2E_j} \qquad (j=0,s,e).
$$

The rate of the discussed reaction is determined by the so-called
(spin-averaged) fully differential cross section
(FDCS)~\cite{kouzakov02},
\begin{eqnarray}
\label{FDCS}\frac{d\sigma}{dE_sd\Omega_sdE_ed\Omega_e}&=&\frac{k_sk_e}{(2\pi)^5k_0}\sum_{i_{\rm
occ}}\left\{\frac14|\langle{\bf k}_s{\bf k}_e|T|{\bf
k}_0i\rangle+\langle{\bf k}_e{\bf k}_s|T|{\bf
k}_0i\rangle|^2\right.\nonumber\\&{}&\left.+\frac34|\langle{\bf
k}_s{\bf k}_e|T|{\bf k}_0i\rangle-\langle{\bf k}_e{\bf k}_s|T|{\bf
k}_0i\rangle|^2\right\} 
\delta(E_s+E_e-E_0-\epsilon_i).
\end{eqnarray}
Here we specified the directions of the momenta of the emitted
electrons by the solid angles $\Omega_{s/e}$. The state vectors
$|{\bf k}_s{\bf k}_e\rangle$ and $|{\bf k}_0i\rangle$ describe,
respectively, the two final-state electrons with asymptotic
momenta ${\bf k}_s$, ${\bf k}_e$ and the initial state consisting
of the projectile state with momentum ${\bf k}_0$ and the
valence-band state $|i\rangle$. The sum is taken over all occupied
one-electron states of the surface with energy
$\epsilon_i=E_s+E_e-E_0$. The operator $T$ is an effective
transition operator that induces the $(e,2e)$ process and is
assumed to be spin independent. In the frozen-core approximation
it has the formal structure
\begin{equation}
\label{trans_operator}
T=V_s+W+(V_s+V_e+W)G^+_{se}(E_{tot})(V_s+W),
\end{equation}
where $V_s$, $V_e$, and $W$ are effective (in general, optical)
electron-solid and electron-electron potentials, respectively, and
$G^+_{se}(E_{tot})$ is the retarded two-electron propagator in the
potential $V_s+V_e+W$ at the total energy $E_{tot}=E_s+E_e$. The
latter satisfies the Lippmann-Schwinger equation
\begin{equation}
\label{green_tot}
G_{se}^+(E_{tot})=G_0^+(E_{tot})+G_0^+(E_{tot})(V_s+V_e+W)G_{se}^+(E_{tot}),
\end{equation}
with $G_0^+(E_{tot})$ being the free two-electron propagator.

%
%
In what follows, we treat Eq.~(\ref{trans_operator}) only to the
first order in the electron-electron interaction $W$. Such a
procedure is usually justified by the choice of the kinematics
such that $E_0\gg\Delta E$ $(\Delta E=E_0-E_s)$ as well as by the
screening of the electron-electron interaction due to the
surrounding medium. This then amounts to the distorted wave Born
approximation (DWBA)
\begin{equation}
\label{DWBA}T=\left[1+(V_s+V_e)G^+_{s,e}(E_{tot})\right]W[1+G^+_{s,e}(E_{tot})V_s],
\end{equation}
where $G^+_{s,e}(E_{tot})$ is the two-electron propagator in the
potential $V_s+V_e$ and is given by the solution of the following
Lippmann-Schwinger equation:
\begin{equation}
\label{green_tot_1p}
G_{s,e}^+(E_{tot})=G_0^+(E_{tot})+G_0^+(E_{tot})(V_s+V_e)G_{s,e}^+(E_{tot}).
\end{equation}

Taking into account Eq.~(\ref{DWBA}), we can present
FDCS~(\ref{FDCS}) as
\begin{eqnarray}
\label{FDCS1}\frac{d\sigma}{d\epsilon_sd\Omega_sd\epsilon_ed\Omega_e}&=&\frac{k_sk_e}{(2\pi)^5k_0}\sum_{i_{\rm
occ}}\left\{\frac14|\langle\chi^{(-)}_{{\bf k}_s}\chi^{(-)}_{{\bf
k}_e}|W|\chi^{(+)}_{{\bf k}_0}i\rangle+\langle\chi^{(-)}_{{\bf
k}_e}\chi^{(-)}_{{\bf k}_s}|W|\chi^{(+)}_{{\bf
k}_0}i\rangle|^2\right.\nonumber\\&{}&\left.+\frac34|\langle\chi^{(-)}_{{\bf
k}_s}\chi^{(-)}_{{\bf k}_e}|W|\chi^{(+)}_{{\bf
k}_0}i\rangle-\langle\chi^{(-)}_{{\bf k}_e}\chi^{(-)}_{{\bf
k}_s}|W|\chi^{(+)}_{{\bf k}_0}i\rangle|^2\right\}\nonumber\\
&{}&\times\delta(E_s+E_e-E_0-\epsilon_i),
\end{eqnarray}
where
\begin{eqnarray}
\label{LEED} \chi^{(+)}_{{\bf k}_0}({\bf r})&=&e^{i{\bf k}_0{\bf
r}}+\int d{\bf r}'e^{i{\bf k}_0{\bf r}'}\upsilon({\bf
r'})g^{+}({\bf r'},{\bf r};E_0),\\ \label{tr_LEED}
\chi^{(-)}_{{\bf k}_j}({\bf r})&=&e^{i{\bf k}_j{\bf r}}+\int d{\bf
r}'e^{i{\bf k}_j{\bf r}'}\upsilon({\bf r'})g^-({\bf r'},{\bf
r};E_j),
\end{eqnarray}
with $j=s,e$ and $V_j\equiv\upsilon$, and $g^+$ ($g^-$) being the
retarded (advanced) one-electron Green's function in the potential
$\upsilon$. In the case of solids with a translational symmetry
parallel to the surface, the states $\chi^{(\pm)}_{{\bf k}}$ can
be calculated within the dynamical low-energy electron diffraction
(LEED) theory~\cite{leed,luders01}.

Employing the surface dielectric function $\varepsilon$ for the
description of the effect of the  screening on the bare
electron-electron interaction $\upsilon_{ee}$, we get
\begin{equation}
\label{ee} W({\bf r},{\bf r}';\omega)=\int\,d{\bf
r}''\varepsilon^{-1}({\bf r},{\bf r}'';\omega)\upsilon_{ee}({\bf
r}'',{\bf r}'), \qquad \upsilon_{ee}({\bf r}'',{\bf
r}')=\frac{1}{|{\bf r}''-{\bf r}'|},
\end{equation}
where $\varepsilon^{-1}$ is the inverse dielectric function, and
the energy argument $\omega$ ($\omega=E_0-E_s$ or
$\omega=E_0-E_e$, depending on the final state of the incident
electron) accounts for the dynamical screening effects.

\section{Surface dielectric response}
\label{surf_resp}
For our purposes we need the inverse dielectric function
$\varepsilon^{-1}$ that is derived from the dielectric function
$\varepsilon$ according to
\begin{equation}
\int d{\bf r}''\varepsilon({\bf r},{\bf
r}'';\omega)\varepsilon^{-1}({\bf r}'',{\bf r}';\omega)=\int d{\bf
r}''\varepsilon^{-1}({\bf r},{\bf r}'';\omega)\varepsilon({\bf
r}'',{\bf r}';\omega)=\delta({\bf r}-{\bf r}').
\end{equation}
Let us assume that the $z$ axis is perpendicular to the surface
and that the solid sample fills the $z<0$ region (see
Fig.~\ref{setup}). If the sample is crystalline, then
\begin{equation}
\varepsilon({\bf r},{\bf r}';\omega)=\varepsilon({\bf r}+{\bf
A},{\bf r}'+{\bf A};\omega), \qquad \varepsilon^{-1}({\bf r},{\bf
r}';\omega)=\varepsilon^{-1}({\bf r}+{\bf A},{\bf r}'+{\bf
A};\omega),
\end{equation}
where ${\bf A}$ is the lattice vector parallel to the surface.
Thus, $\varepsilon$ and $\varepsilon^{-1}$ can be presented as
\begin{eqnarray}
\varepsilon({\bf r},{\bf r}';\omega) &=& \sum_{{\bf G},{\bf
G}'}e^{i({\bf G}{\bf R}-{\bf G}'{\bf R}')}\int\limits_{\rm
1^{st}\, BZ}\frac{d{\bf Q}}{(2\pi)^2}\,\varepsilon_{{\bf G},{\bf
G}'}({\bf
Q},z,z';\omega)e^{i{\bf Q}({\bf R}-{\bf R}')},\\
\varepsilon^{-1}({\bf r},{\bf r}';\omega) &=& \sum_{{\bf G},{\bf
G}'}e^{i({\bf G}{\bf R}-{\bf G}'{\bf R}')}\int\limits_{\rm
1^{st}\, BZ}\frac{d{\bf Q}}{(2\pi)^2}\,\varepsilon_{{\bf G},{\bf
G}'}^{-1}({\bf Q},z,z';\omega)e^{i{\bf Q}({\bf R}-{\bf
R}')},\label{inverse}
\end{eqnarray}
where ${\bf r}=({\bf R},z)$ and ${\bf r}'=({\bf R}',z')$, ${\bf
G}$ and ${\bf G}'$ are the surface reciprocal lattice vectors, and
the ${\bf Q}$ integration is carried over the first Brillouin zone
($1^{\rm st}$\,BZ) of the surface reciprocal lattice. Using
Eq.~(\ref{inverse}), we get for the screened Coulomb
potential~(\ref{ee})
\begin{eqnarray}\label{crystal}
W({\bf r},{\bf r}_0;\omega)&=&\sum_{{\bf G},{\bf G}'}e^{i({\bf
G}{\bf R}-{\bf G}'{\bf R}_0)}\int\limits_{\rm 1^{st}\,
BZ}\frac{d{\bf Q}}{2\pi}\,\frac{e^{i{\bf Q}({\bf R}-{\bf
R}_0)}}{|{\bf Q}+{\bf G}'|}
\int dz'\varepsilon_{{\bf G},{\bf G}'}^{-1}({\bf
Q},z,z';\omega)e^{-|{\bf Q}+{\bf G}'||z'-z_0|}.\nonumber\\
\end{eqnarray}
If we neglect the crystalline effects on $\varepsilon$ and
$\varepsilon^{-1}$ parallel to the surface, then
\begin{equation}
\varepsilon^{-1}_{{\bf G},{\bf G}'}({\bf
Q},z,z';\omega)=\delta_{{\bf G},{\bf G}'}\varepsilon^{-1}({\bf
Q}+{\bf G},z,z';\omega),
\end{equation}
where it is assumed that ${\bf Q}\in1^{\rm st}$~BZ, and hence
%
\begin{equation}\label{nocrystal}
W({\bf r},{\bf r}_0;\omega)=\int\frac{d{\bf
Q}}{2\pi}\,\frac{e^{i{\bf Q}({\bf R}-{\bf R}_0)}}{Q}\int
dz'\varepsilon^{-1}({\bf Q},z,z';\omega)e^{-Q|z'-z_0|}.
\end{equation}

Representations~(\ref{crystal}) and~(\ref{nocrystal}) can be
particularly useful, when employing the following expansion of the
electron states in Eq.~(\ref{FDCS1}):
\begin{equation}
\chi^{(\pm)}_{\bf k}({\bf r})=\sum_{\bf G}C_{\bf k}^{(\pm)}({\bf
G},z)e^{i({\bf K}+{\bf G}){\bf R}},
\end{equation}
where ${\bf k}=({\bf K},k_z)$.

If the sample is a free-electron metal (for instance, Al or Be),
then the model of a degenerate electron gas where electrons move
in a positive ionic background and are bounded by a surface
potential barrier is commonly applicable to mimic its dielectric
properties. Below, we briefly sketch two possible approaches for
describing the dielectric response of such  jellium-like systems.

\subsection{Specular-reflection model}
\label{SRM}
The problems involving the interaction of  charged particles with
plane-bounded solids are often treated using SRM. This model
assumes the surface barrier to be impermeable for the electrons in the
solid, so that they are specularly reflected at the surface. The
reflection process is described in a classical
fashion, in particular the interference between the incoming and outgoing components
is neglected. Within SRM the potential created by the external
charge distribution $\rho({\bf r},t)$ in the vicinity of a surface
is given by (see, for instance, Ref.~\cite{echenique92} and
references therein)
\begin{equation}\label{srm_pot}
V({\bf r},t)=V_+({\bf r},t)\Theta(z)+V_-({\bf r},t)\Theta(-z),
\qquad V_\pm({\bf r},t)=\int\frac{d{\bf
Q}}{(2\pi)^2}\int\frac{d\omega}{2\pi}\,V_\pm({\bf
Q},z;\omega)e^{i({\bf QR}-\omega t)},
\end{equation}
where
\begin{equation}\label{srm_pot_1}
V_\pm({\bf Q},z;\omega)=4\pi[U_\pm({\bf
Q},z;\omega)\mp\tilde\rho({\bf Q},\omega)\nu_\pm({\bf
Q},z;\omega)],
\end{equation}
with [using the notation ${\bf q}=({\bf Q},q_z)$]
\begin{eqnarray}\label{srm_pot_2}U_+({\bf
Q},z;\omega)=\int\frac{dq_z}{2\pi}\frac{\rho_+({\bf
q},\omega)}{q^2}\,e^{iq_zz}, \qquad U_-({\bf
Q},z;\omega)=\int\frac{dq_z}{2\pi}\frac{\rho_-({\bf
q},\omega)}{q^2\varepsilon_b({\bf q},\omega)}\,e^{iq_zz},
\end{eqnarray}
$$
\rho_\pm({\bf q},\omega)=\int d{\bf r}\int dt\,\rho({\bf
R},\pm|z|,t)e^{-i({\bf qr}-\omega t)},
$$
\begin{eqnarray}\label{srm_pot_3}\nu_+({\bf
Q},z;\omega)=\int\frac{dq_z}{2\pi}\frac{2Q}{q^2}\,e^{iq_zz}=e^{-Q|z|},
\qquad \nu_-({\bf
Q},z;\omega)=\int\frac{dq_z}{2\pi}\frac{2Q}{q^2\varepsilon_b({\bf
q},\omega)}\,e^{iq_zz},
\end{eqnarray}
\begin{equation}\label{srm_pot_4}\tilde\rho({\bf
Q},\omega)=\frac{1}{1+\varepsilon_s({\bf Q},\omega)}[U_+({\bf
Q},0;\omega)-U_-({\bf Q},0;\omega)], \qquad \varepsilon_s({\bf
Q},\omega)=\nu_-({\bf Q},0;\omega).
\end{equation}
In Eqs.~(\ref{srm_pot_2}) and~(\ref{srm_pot_3}),
$\varepsilon_b({\bf q},\omega)$ is the bulk dielectric function,
i.e., that of an infinite 3D system. The quantity
$\varepsilon_s({\bf Q},\omega)$ occurring in Eq.~(\ref{srm_pot_4})
is the so-called surface dielectric function~\cite{newns70}.

To utilize the SRM approach in the present quantum-mechanical
treatment, the external charge density $\rho({\bf r},t)$ must be
replaced by the corresponding operator $\rho({\bf
r},t)=e^{-iH_0t}\rho({\bf r})e^{iH_0t}$, where $H_0$ is the
Hamiltonian associated with the incident electron. Clearly, in
this way the frequency $\omega$ equals the energy transfer $\Delta
E=E_0-E_{s(e)}$. Since $\rho({\bf r})=\delta({\bf r}-{\bf r}_0)$,
where ${\bf r}_0$ is the position of the incident electron, it is
straightforward to show that the resulting expression for the
inverse dielectric function in Eq.~(\ref{nocrystal}) is
\begin{eqnarray}\label{srm}
\varepsilon^{-1}({\bf
Q},z,z';\omega)&=&\Theta(z)\left\{\delta(z-|z'|)\Theta(z_0)-\frac{2e^{-Qz}}{1+\varepsilon_s({\bf
Q},\omega)}[\delta(z')\Theta(z_0)- \kappa({\bf
Q},z';\omega)\Theta(-z_0)]\right\}\nonumber\\&{}&+\Theta(-z)\Bigg\{[\kappa({\bf
Q},z+z';\omega)+\kappa({\bf
Q},z-z';\omega)]\Theta(-z_0)+\frac{2\nu_-({\bf
Q},z;\omega)}{1+\varepsilon_s({\bf
Q},\omega)}\nonumber\\&{}&\times[\delta(z')\Theta(z_0)-
\kappa({\bf Q},z';\omega)\Theta(-z_0)]\Bigg\},
\end{eqnarray}
where
\begin{eqnarray}\label{srm_funcs}
\kappa({\bf
Q},z;\omega)=\int\limits^\infty_{-\infty}\frac{dq_z}{2\pi}\frac{e^{iq_zz}}{\varepsilon_b({\bf
q},\omega)}.
\end{eqnarray}
Note that the inverse dielectric function~(\ref{srm}) depends on
whether the incoming electron is inside ($z_0<0$) or outside
($z_0>0$) the solid.

\subsection{Random phase approximation}
\label{IBM}
RPA constitutes a reasonable framework for describing the
dielectric response of a degenerate electron gas. The bulk
dielectric function $\varepsilon_b$ was first derived within this
method by Lindhard~\cite{lindhard}. For the case of  a
semi-infinite geometry a very useful study is from
Newns~\cite{newns70}, who considered the dielectric response of a
semi-infinite ideal metal within RPA assuming an infinite surface
barrier (RPA-IB), which is in the spirit of SRM. Using the
approximation of a specular electron reflection, Bechstedt
\emph{et al.}~\cite{bechstedt83} derived expressions for the
screened Coulomb potential and the inverse dielectric function.
However, Horing \emph{et al.}~\cite{horring85} noticed that the
result of Ref.~\cite{bechstedt83} is incapable of describing
correctly the image field as a part of the dynamically screened
Coulomb potential. In their calculation, Horing \emph{et al.}
utilized the potential solutions obtained by Newns~\cite{newns70}
within the RPA-IB model. Employing the mathematical model
delta-function potential as a bare, unscreened interaction and
neglecting the nondiagonal elements in the density-response
matrix, they calculated the inverse dielectric function as
\begin{eqnarray}\label{newns}
\varepsilon^{-1}({\bf
Q},z,z';\omega)&=&\Theta(z)\left\{\delta(z-z')-\frac{e^{-Qz}}{1+\varepsilon_s({\bf
Q},\omega)}[\delta(z')- 2\kappa({\bf
Q},z';\omega)\Theta(-z')]\right\}\nonumber\\&{}&+\Theta(-z)\Bigg\{[\kappa({\bf
Q},z+z';\omega)+\kappa({\bf
Q},z-z';\omega)]\Theta(-z')+\frac{\nu_-({\bf
Q},z;\omega)}{1+\varepsilon_s({\bf
Q},\omega)}\nonumber\\&{}&\times[\delta(z')-2\kappa({\bf
Q},z';\omega)\Theta(-z')]\Bigg\},
\end{eqnarray}
where the functions $\kappa({\bf Q},z;\omega)$, $\nu_-({\bf
Q},z;\omega)$, and $\varepsilon_s({\bf Q},\omega)$ are the same as
in Sec.~\ref{SRM}.

Expressions~(\ref{srm}) and~(\ref{newns}) have a similar
structure. Moreover, one can formally deduce Eq.~(\ref{newns})
using the SRM approach in the case of the model delta-function
potential $\upsilon_{ee}({\bf r},{\bf r}_0)=\delta({\bf r}-{\bf
r}_0)$. However, the RPA-IB result~(\ref{newns}) is supposed to be
applicable in the case of Coulomb-like potentials as well (see
Ref.~\cite{horring85} for detail). The latter feature makes the
two approaches nonequivalent.

\section{Results and discussion}
\label{res}
In this section we present and analyze the numerical results for
the correlated electron-pair emission from  Al and Be surfaces. To
be specific, we consider the kinematics of the recent experimental
study on Al~\cite{werner08}. In that experiment, the electron
energy loss was measured at the incident energy $E_0=100$\,eV and
the angle $\theta_0=30^\circ$ in the specular reflection mode,
that is, $\theta_s=30^\circ$, in coincidence with the secondary
electron ejected at the angle $\theta_e=60^\circ$ (see
Fig.~\ref{setup}). We neglect the crystalline effects in our
calculations and construct the one-electron states in
Eq.~(\ref{FDCS1}) in the context of the jellium model. The details
of this procedure are given in the Appendix. The surface
dielectric response is taken into account within the SRM and
RPA-IB approaches described in the previous section. Each of them
depends on the model of the bulk dielectric response. Below we
outline two approximations for the bulk dielectric function of a
free-electron metal that were employed in the present
calculations: (i) the Thomas-Fermi (TF)~\cite{ashcroft_book} and
(ii) the hydrodynamic (HA)~\cite{bloch33} approximations. In spite
of their relative simplicity, they efficiently mimic the basic
features pertinent to the static and dynamical bulk screening
effects in metals. In addition, their use in the calculations
makes the numerical implementation more transparent and
controllable.

\subsubsection{Thomas-Fermi approximation}
This well-known model neglects the  dynamical screening effects
and assumes the dielectric function to be of the form
\begin{equation}\label{TF}
\varepsilon_b(q,\omega)=1+\frac{\lambda^2}{q^2},
\end{equation}
where $\lambda$ is the screening constant. For a free-electron
isotropic gas one has
$$
\lambda^2=\frac{4k_F}{\pi}, \qquad k_F=\sqrt{2\epsilon_F},
$$
where $\epsilon_F$ and $k_F$ are the Fermi energy and momentum,
respectively. Using the Wigner-Seitz radius $r_s$, we have
$$
\lambda^2=\left(\frac{12}{\pi}\right)^{2/3}\frac{1}{r_s}
$$
($r_s=2.07$ for Al and $r_s=1.87$ for Be).

Within the TF model the functions in Eqs.~(\ref{srm})
and~(\ref{newns}) are given by
\begin{eqnarray}\label{newns_funcs_TF}
\kappa({\bf
Q},z;\omega)=\delta(z)-\frac{\lambda^2}{2\Lambda}\,e^{-\Lambda|z|},
\qquad \nu_-({\bf Q},z;\omega)=\frac{Q}{\Lambda}\,e^{-\Lambda|z|},
\qquad \varepsilon_s({\bf
Q},\omega)=\frac{Q}{\Lambda},
\end{eqnarray}
where $\Lambda=\sqrt{Q^2+\lambda^2}$.

\subsubsection{Hydrodynamic approximation}
Since its introduction almost 80 years ago~\cite{bloch33}, the
hydrodynamic approximation has proved to be very useful in
describing the electrical transport and the optical properties of
conductors. The main advantage of this model lies in the
simplicity of accounting for the spatial dispersion,
\begin{equation}\label{HA}
\varepsilon_b(q,\omega)=1+\frac{\omega_{b}^2}{\beta^2q^2-\omega(\omega+i\nu)},
\end{equation}
where the frequency of the bulk plasmon mode $\omega_{b}$ is given
by
$$
\omega_{b}=\sqrt{4\pi n}=\sqrt{\frac{3}{r_s^3}},
$$
with $n$ being the electron density. The parameter $\beta$ depends
on $\omega$, such that
\begin{equation}
\beta=\left\{\begin{array}{l}\displaystyle\sqrt{\frac{1}{3}}\upsilon_F,\qquad
\omega\ll\nu\\
\displaystyle\sqrt{\frac{3}{5}}\upsilon_F,\qquad
\omega\gg\nu.\end{array}\right.
\end{equation}
The above low- and high-frequency limits can be reproduced with
the ``interpolation formula''~\cite{halevi95}
\begin{equation}
\beta^2=\frac{\frac35\omega+\frac13i\nu}{\omega+i\nu}\upsilon_F^2.
\end{equation}
The collision frequency $\nu$ can be estimated as
$\nu\sim\Gamma_{pl}$ ($\Gamma_{pl}=0.53$\,eV for Al and
$\Gamma_{pl}=4.7$\,eV for Be). Note that the TF result~(\ref{TF})
derives from Eq.~(\ref{HA}) in the limit $\omega\to0$.

Within the HA model the functions in Eqs.~(\ref{srm})
and~(\ref{newns}) are given by
\begin{eqnarray}\label{newns_funcs_HA}
\kappa({\bf
Q},z;\omega)&=&\delta(z)-\frac{\omega_{b}^2}{2\beta^2}\,\frac{e^{-\Lambda|z|}}
{\Lambda},\nonumber\\
\qquad \nu_-({\bf
Q},z;\omega)&=&\frac{\omega(\omega+i\nu)e^{-Q|z|}}{\omega(\omega+i\nu)-\omega_{b}^2}
-\frac{Q}{\Lambda}\,\frac{\omega_{b}^2e^{-\Lambda|z|}}{\omega(\omega+i\nu)-\omega_{b}^2},\nonumber\\
\qquad \varepsilon_s({\bf
Q},\omega)&=&\frac{\omega(\omega+i\nu)\Lambda-\omega_{b}^2Q}{[\omega(\omega+i\nu)-\omega_{b}^2]\Lambda},
\end{eqnarray}
where
$\Lambda=-i\beta^{-1}\sqrt{\omega(\omega+i\nu)-\omega_{b}^2-\beta^2Q^2}$
(here the first branch of the square root function is assumed,
which yields ${\rm Re}(\Lambda)>0$).

\subsection{Aluminum}
\label{Al}
\begin{figure}
\begin{center}
\includegraphics[scale=0.75]{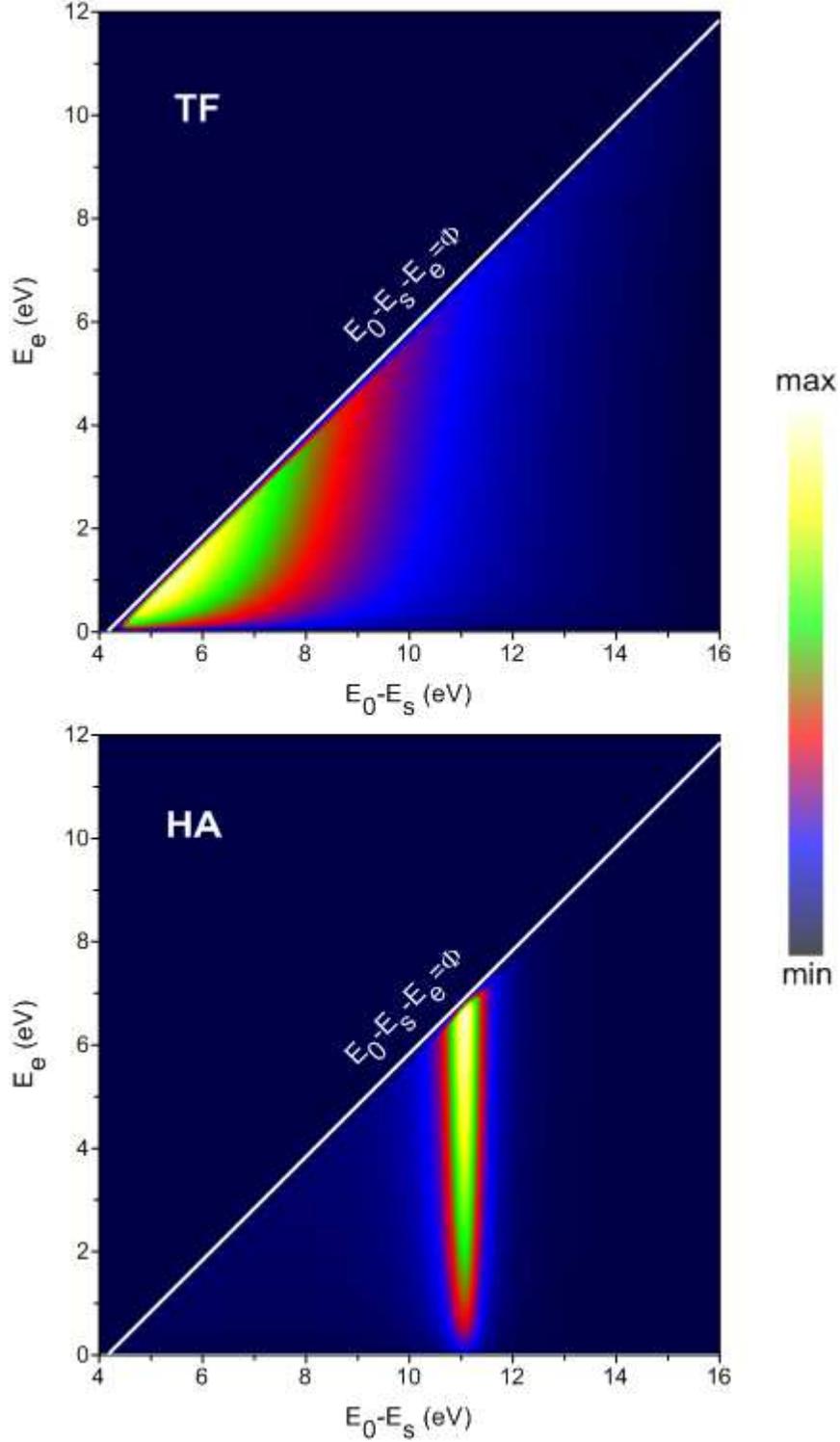}
\end{center}
\caption{\label{srm_Al}(Color online) Correlated electron energy
distribution from Al within SRM. The top panel corresponds to the
TF bulk dielectric function, while the bottom corresponds to HA.
The white solid line marks the energy threshold for the
electron-pair emission, $E_0-E_s-E_e\geq\Phi$, where $\Phi$ is the
work function.}
\end{figure}

Fig.~\ref{srm_Al} shows the correlated electron energy
distribution of the scattered and the ejected electrons calculated
using the SRM surface dielectric function~(\ref{srm}). Marked
differences between the TF and HA models can be seen. The maximum
of the intensity in the TF case is located at small values of
$E_e\sim1$\,eV when the energy loss $\Delta E=E_0-E_s$ exceeds the
work function $\Phi=4.3$\,eV by approximately the same amount,
which indicates that the ejected electron originates from the
initial state close to the Fermi level $\epsilon_F$. With
increasing $\Delta E$ the intensity decreases and a tendency can
be observed: at a given energy-loss value the electrons are
preferably ejected with energies close to the threshold
$E_e=\Delta E-\Phi$, which corresponds to $\epsilon_i=\epsilon_F$.
The latter is readily explained by the fact that the density of
states at the Fermi level is maximal. In contrast to the TF model,
in the HA case the intensity is strongly peaked around $\Delta
E\approx11$\,eV which is slightly above the surface plasmon energy
$\omega_s=\omega_b/\sqrt{2}=10.5$\,eV. The reason for such a
resonant behavior is due to the poles of the
function~$(\ref{srm})$ when $1+\varepsilon_s=0$. If neglecting the
plasmon dispersion and the damping effects in the HA
model~(\ref{HA}), that is, $\beta=0$ and $\nu=0$, one finds that,
according to Eq.~(\ref{newns_funcs_HA}), the pole is located
exactly at $\Delta E=\omega_s$. The plasmon dispersion and the
damping effects are responsible for the shift and for the finite
width of the observed resonant peak.

\begin{figure}
\begin{center}
\includegraphics[scale=0.75]{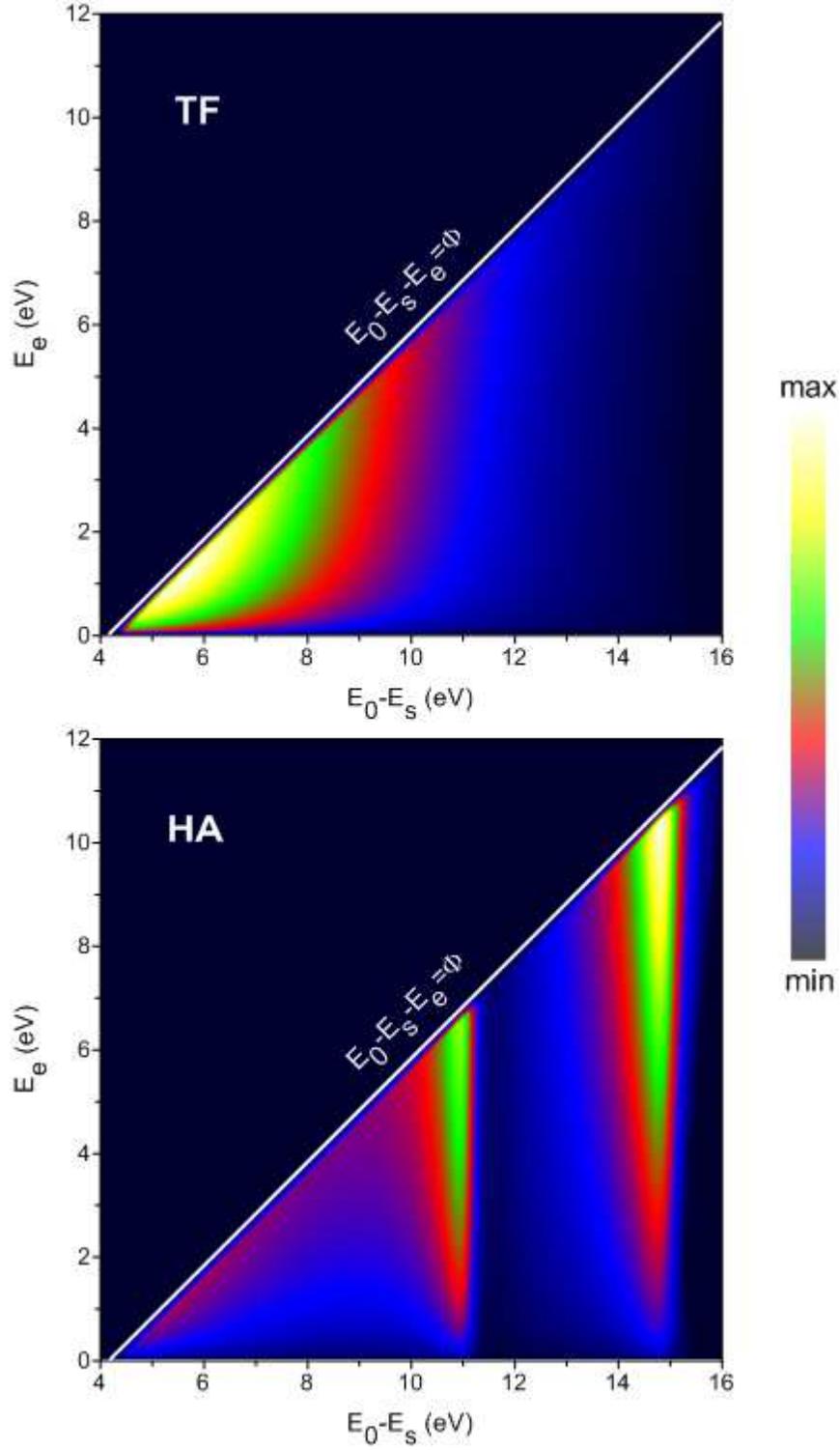}
\end{center}
\caption{\label{newns_Al}(Color online) The same as in
Fig.~\ref{srm_Al}, but using RPA-IB.}
\end{figure}

The numerical results for the correlated electron energy
distribution using the RPA-IB surface dielectric
function~(\ref{newns}) are presented in Fig.~\ref{newns_Al}. The
TF distribution is very close to that when using SRM. However, in
the HA case the distribution differs from the analogous one in
Fig.~\ref{srm_Al}. Namely, in addition to the peak associated with
the surface plasmon there appears another, more pronounced feature
at approximately $\Delta E=15$\,eV. Taking into account that
$\omega_b=14.9$\,eV, it is clear that the feature is due to the
dynamical screening effects related to the bulk-plasmon mode. The
absence of the bulk-plasmon peak in the SRM case and its
appearance in the RPA-IB case follows from the comparison of
Eqs.~(\ref{srm}) and~(\ref{newns}). The bulk screening effects in
these models are associated with function~(\ref{srm_funcs}). Thus,
within the SRM~(\ref{srm}) it comes into play when the incident
electron penetrates inside the metal ($z_0<0$), while within the
RPA-IB~(\ref{newns}) it becomes already relevant when the incident
electron is still moving in the vacuum. A short inelastic mean
free path for the incoming electron results in small penetration
lengths, thus strongly restricting the bulk contribution in the
SRM case.

\begin{figure}
\begin{center}
\includegraphics[scale=0.4]{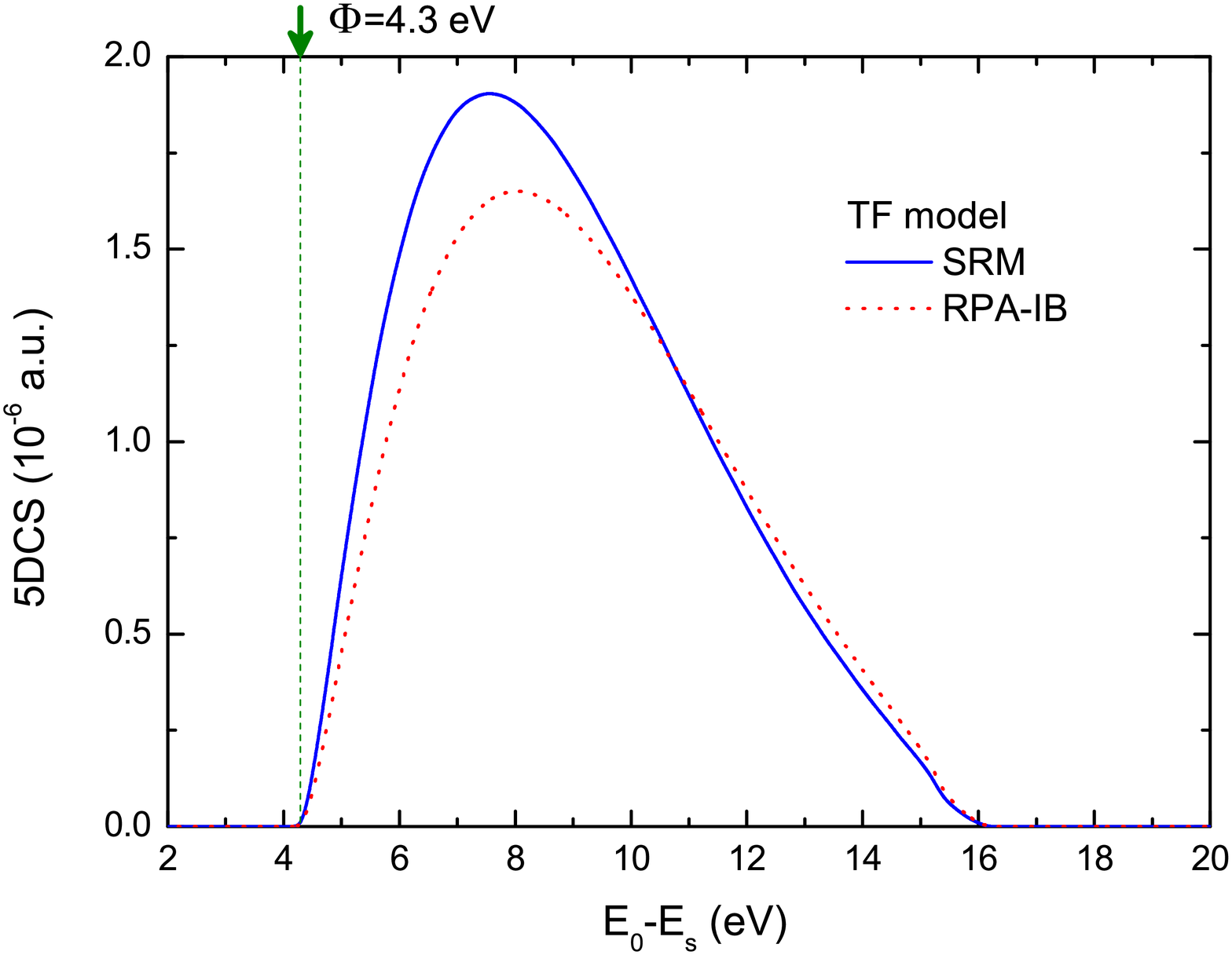}\\[1cm]
\includegraphics[scale=0.4]{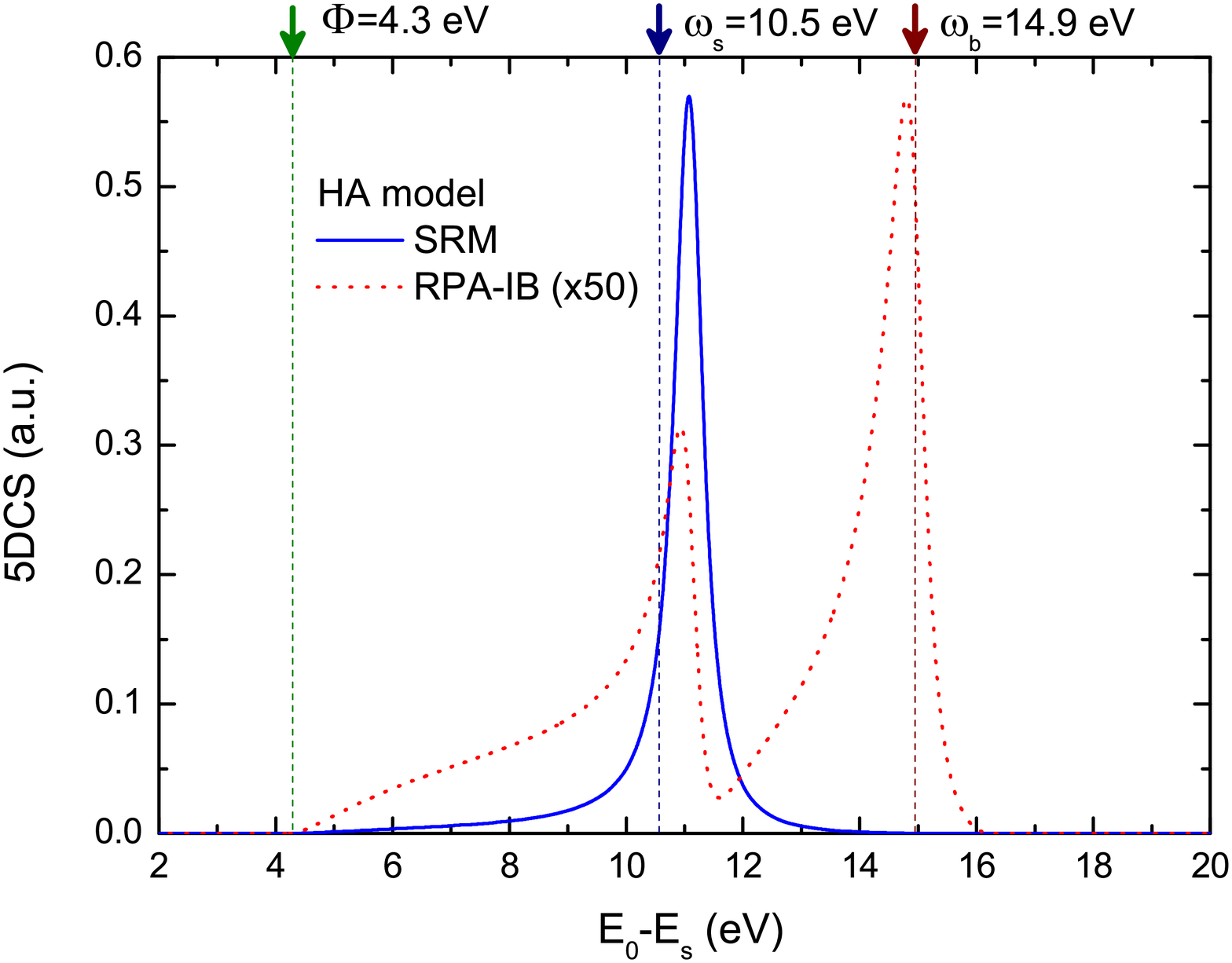}
\end{center}
\caption{\label{int_Al}(Color online) Dependence of the
ejected-electron yields from Al on the energy loss within SRM and
RPA-IB. The top (bottom) panel presents the results using the TF
(HA) bulk dielectric function.}
\end{figure}

Fig.~\ref{int_Al} compares the so-called five-fold differential
cross sections (5DCS), which derive from FDCS~(\ref{FDCS1}) upon
integrating over the ejected electron energy $E_e$, using the SRM
and RPA-IB approaches. Clearly, 5DCS characterizes the dependence
of the ejected-electron yield on the energy loss $\Delta E$ in the
considered collision geometry. In accordance with the TF results
in Figs.~\ref{srm_Al} and~\ref{newns_Al}, the TF ejected-electron
yields in Fig.~\ref{int_Al} are very close to each other, starting
to grow at the threshold value $\Delta E=\Phi$, exhibiting maximum
at $\Delta E=7-8$\,eV, and then smoothly decreasing down to 0 at
$\Delta E\approx16$\,eV, which is due to the restriction $|{\bf
K}_s+{\bf K}_e-{\bf K}_0|\leq k_F$ imposed by the conservation of
the electron-pair momentum parallel to the surface. The HA results
are orders of magnitude larger than the TF ones, indicating the
strength of the dynamical screening effects. In accordance with
Figs.~\ref{srm_Al} and~\ref{newns_Al}, the HA results in
Fig.~\ref{int_Al} using SRM exhibit only one peak associated with
the surface-plasmon mode while those using RPA-IB two peaks which
are due to both the surface- and the bulk-plasmon modes. The
intensity of the surface-plasmon peak in the SRM case exceeds that
in the RPA-IB case by almost two orders of magnitude. The position
and the intensity of the bulk-plasmon peak in the RPA-IB case is
mainly determined by the interplay between the plasmon pole in the
HA bulk dielectric function and the kinematical effects related to
the conservation of the electron-pair energy and the
surface-parallel momentum.

\subsection{Beryllium}
\label{Be}
\begin{figure}
\begin{center}
\includegraphics[scale=0.75]{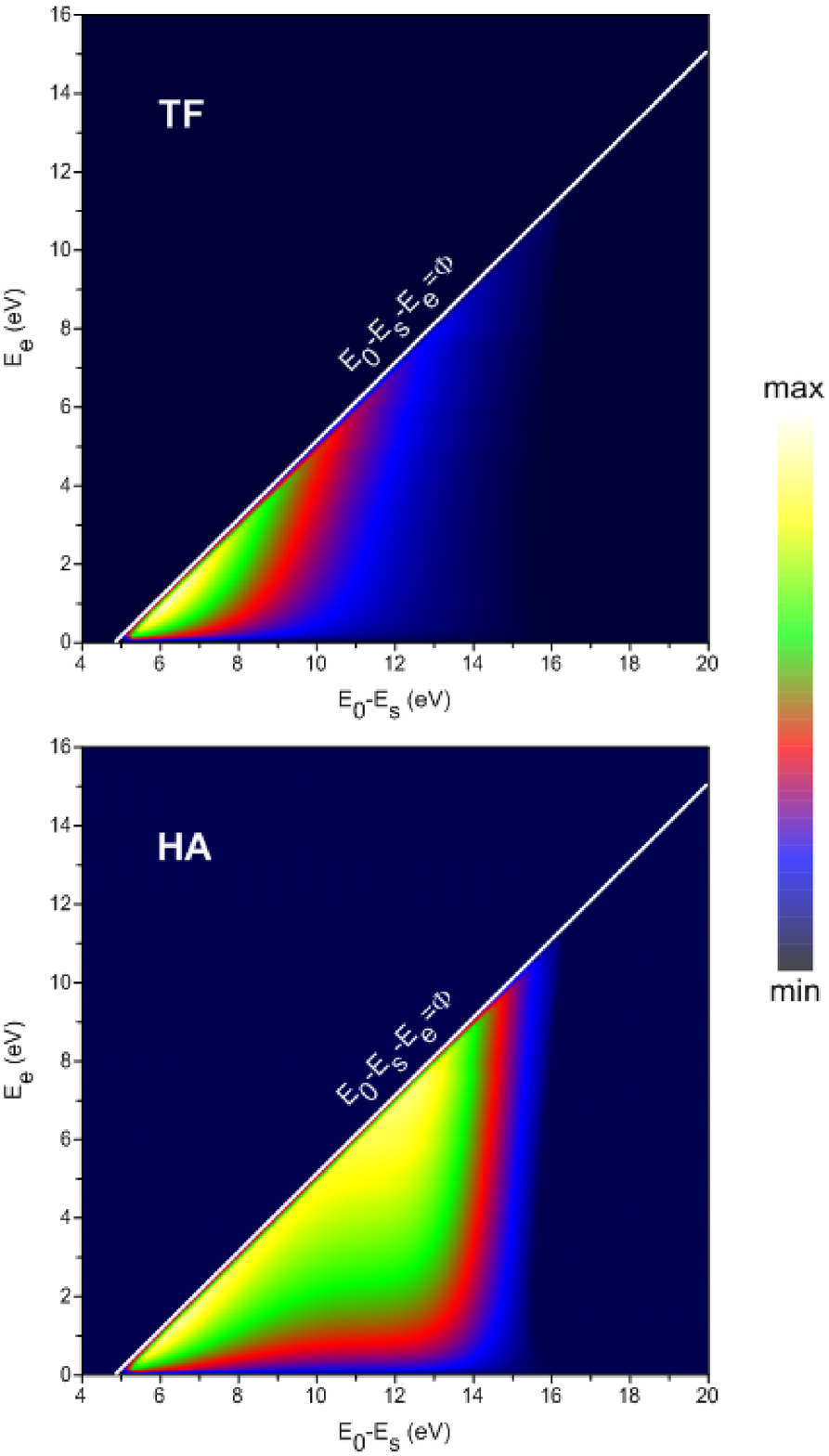}
\end{center}
\caption{\label{srm_Be}(Color online) The same as in
Fig.~\ref{srm_Al}, but in the case of Be.}
\end{figure}

\begin{figure}
\begin{center}
\includegraphics[scale=0.75]{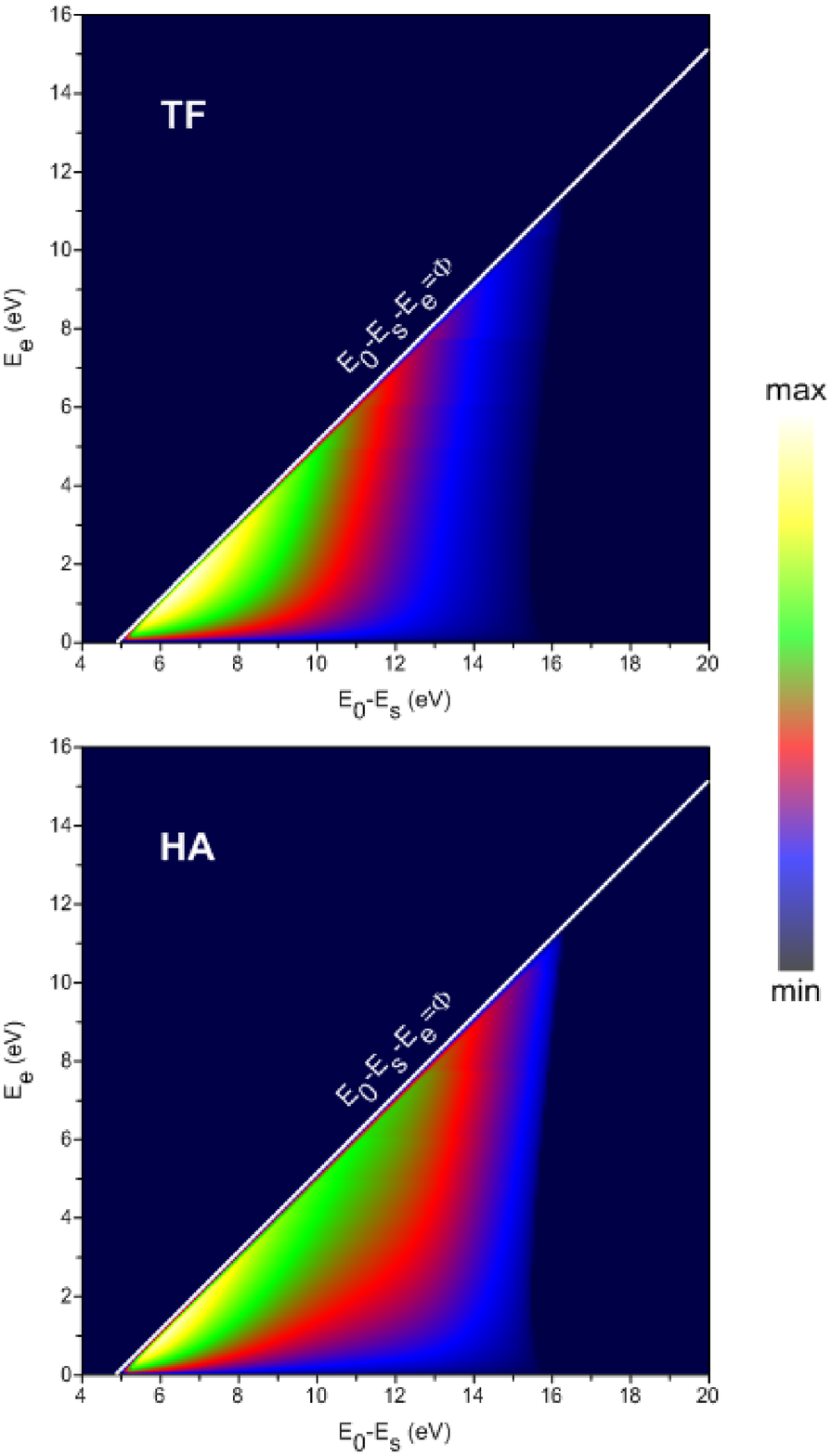}
\end{center}
\caption{\label{newns_Be}(Color online) The same as in
Fig.~\ref{newns_Al}, but in the case of Be.}
\end{figure}

\begin{figure}
\begin{center}
\includegraphics[scale=0.4]{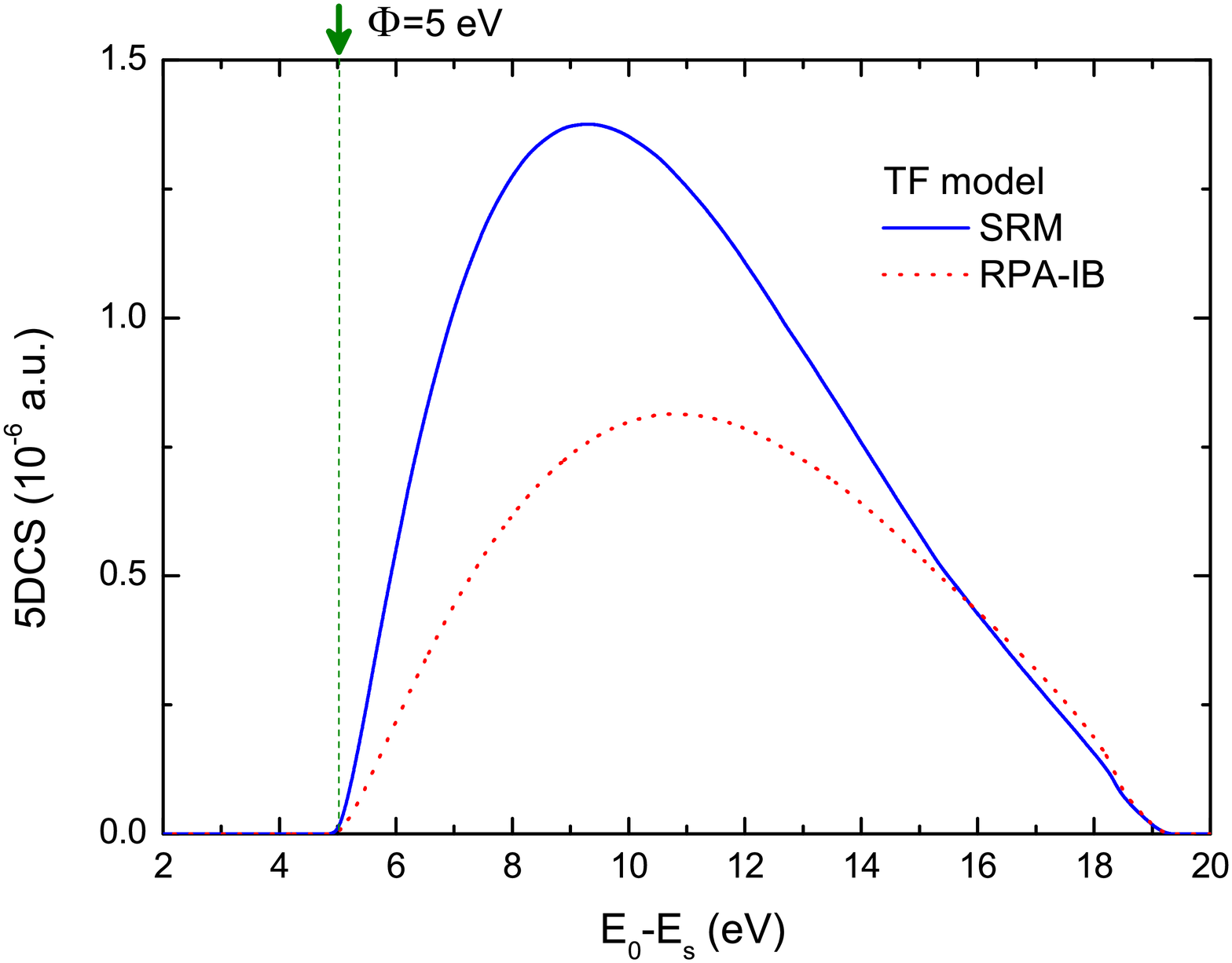}\\[1cm]
\includegraphics[scale=0.4]{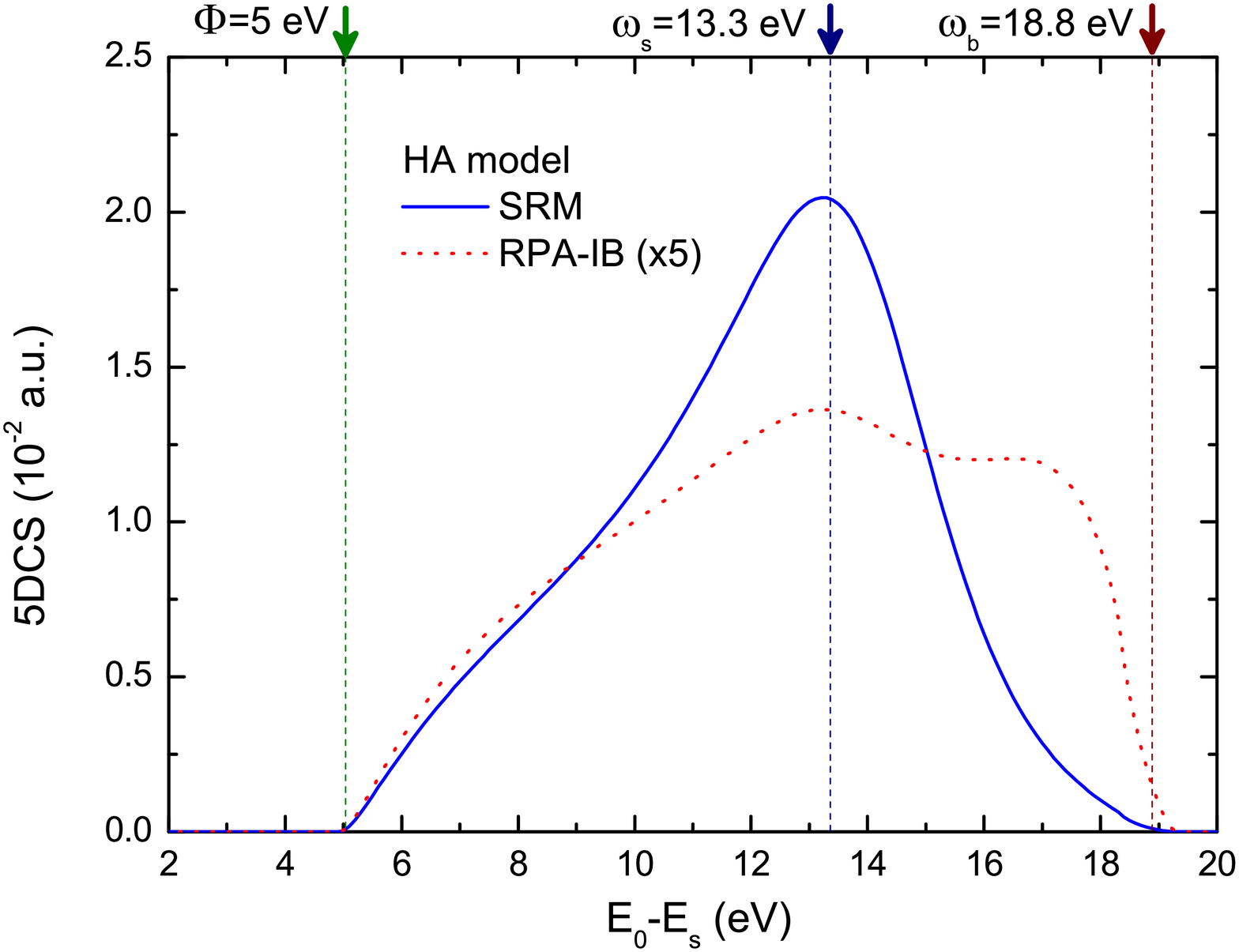}
\end{center}
\caption{\label{int_Be}(Color online) The same as in
Fig.~\ref{int_Al}, but in the case of Be.}
\end{figure}

Figs.~\ref{srm_Be},~\ref{newns_Be}, and~\ref{int_Be} present
numerical calculations for the same situations as in
Figs.~\ref{srm_Al},~\ref{newns_Al}, and~\ref{int_Al},
respectively, but for Be. In the HA case, the differences compared
to the results for Al can be attributed to a much wider plasmon
resonance in Be. In particular, we find that the footprints of the
plasmon modes in the HA results are much weaker and broader. The
role of the plasmon linewidth $\Gamma_{pl}$ is clearly seen when
comparing the HA results for Al and Be using SRM. The FWHM of the
peaks observed in the HA results within SRM in Figs.~\ref{int_Al}
and~\ref{int_Be} are given by approximately $\Gamma_{pl}({\rm
Al})=0.53$\,eV and $\Gamma_{pl}({\rm Be})=4.7$\,eV, respectively.
Moreover, when using RPA-IB, the plasmon features can hardly be
identified in FDCS (Fig.~\ref{newns_Be}). And they manifest
themselves for some cases only in 5DCS (Fig.~\ref{int_Be}). In
contrast, the TF results for Be are rather close to those for Al,
particularly in magnitude. Though the differences between the TF
and HA results in Figs.~\ref{srm_Be} and~\ref{newns_Be} are not as
large as in the case of Al, the effects of dynamical screening are
still quite strong. This conclusion follows from the fact that the
HA results for 5DCS are about four orders of magnitude lager than
the TF ones (see Fig.~\ref{int_Be}).

\section{Theory and experiment}
\label{th&exp}
Here we discuss how the predictions  of the  present theoretical
approach compare to the results of the recent experiments on
Al~\cite{werner08,werner11}. The kinematical regime of
Ref.~\cite{werner08} was specified in the previous section. The
measurements in Ref.~\cite{werner11} were carried out for a normal
incidence ($\theta_0=0^\circ$) of the projectile electron with an
impact energy of 500\,eV. The scattered and the ejected electrons
were detected at the polar angles $\theta_s=\theta_e=60^\circ$
(see Fig.~\ref{setup}). In both studies, the secondary-electron
spectra were recorded in coincidence with the primary electron
having undergone an energy loss that is equal to the bulk- or the
surface-plasmon frequencies, i.e., when $\Delta E=\omega_b$ and
$\Delta E=\omega_s$. In coincident measurements of the
secondary-electron spectra in Ref.~\cite{werner11}, further data
were reported for the energy-loss values of 25, 30, 40, 45, and
150\,eV.

For a  detailed quantitative comparison of our theoretical results
with the discussed measurements one should consider the following
aspects. First, the experimental data are reported  on  an
arbitrary intensity scale, which means that the absolute values of
the FDCS are not determined. Second, one must bear in mind that
the experimental data are broadened by a finite energy and angular
resolution whose  effective values in Ref.~\cite{werner08} were
given to be 1 and 1.2\,eV for the primary (scattered) and the
secondary (ejected) electrons, respectively, while in
Ref.~\cite{werner11} the overall energy resolution was reported to
be about 5\,eV. Below we take into account the energy-broadening
effect by convoluting our theoretical calculations with a Gaussian
energy distribution.

\begin{figure}
\begin{center}
\includegraphics[scale=0.4]{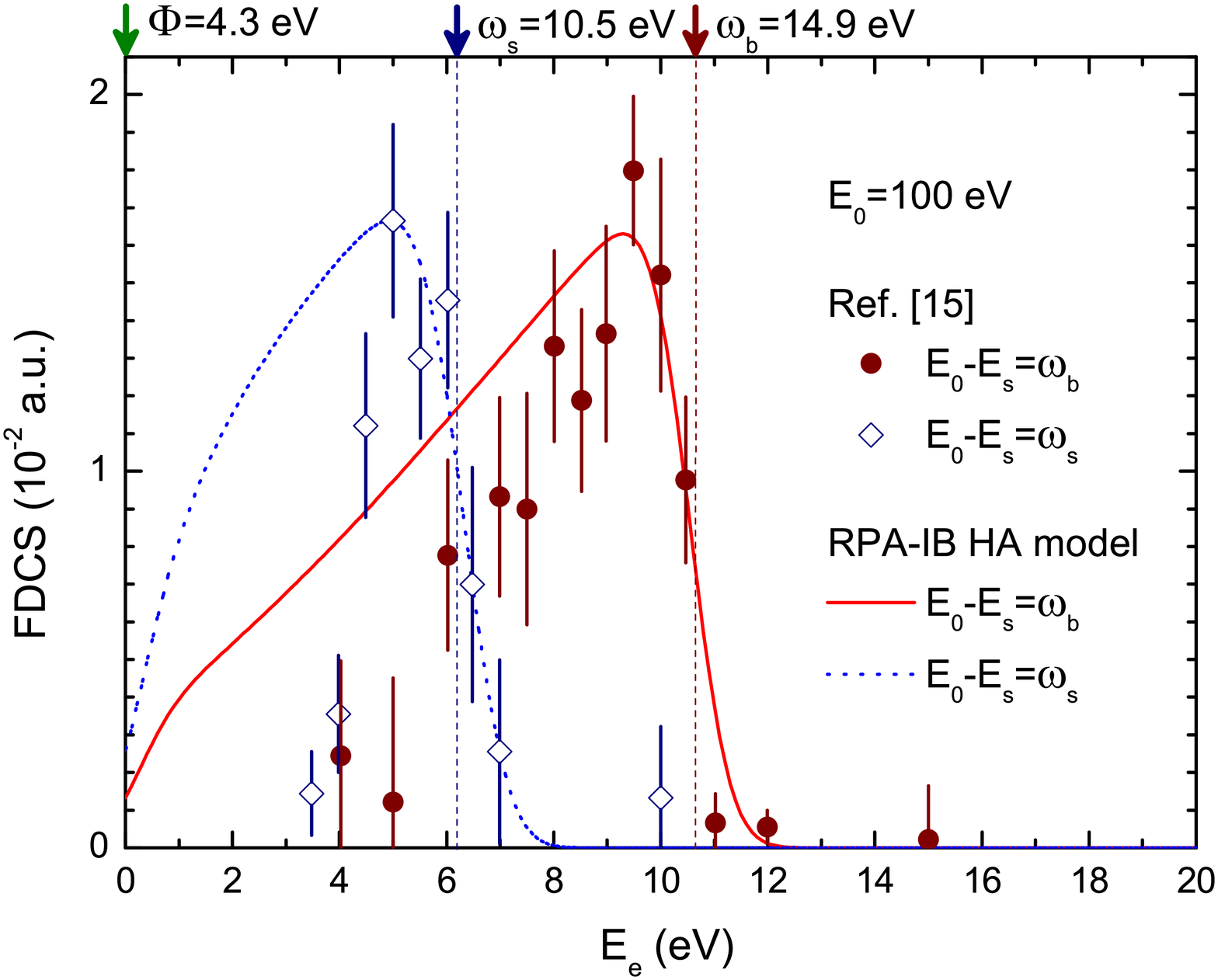}\\[1cm]
\includegraphics[scale=0.4]{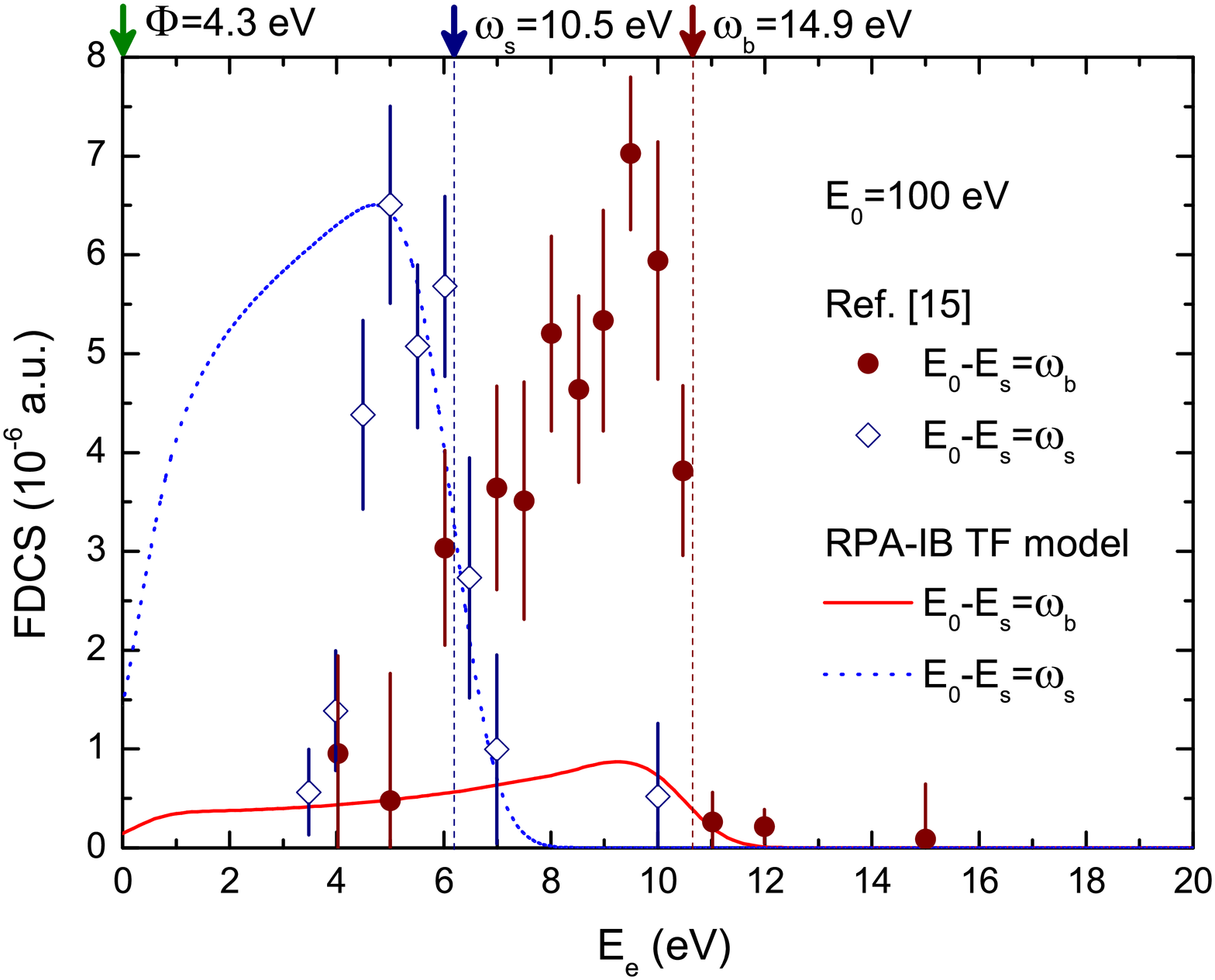}
\end{center}
\caption{\label{Al_100_Newns}(Color online) Spectra of secondary
electrons at energy losses equal to the surface- and bulk-plasmon
frequencies. The calculations were carried out within the RPA-IB
model using the HA (top panel) and TF (bottom panel) bulk
dielectric functions. The experimental data points are borrowed
from Ref.~\cite{werner08}.}
\end{figure}

\begin{figure}
\begin{center}
\includegraphics[scale=0.4]{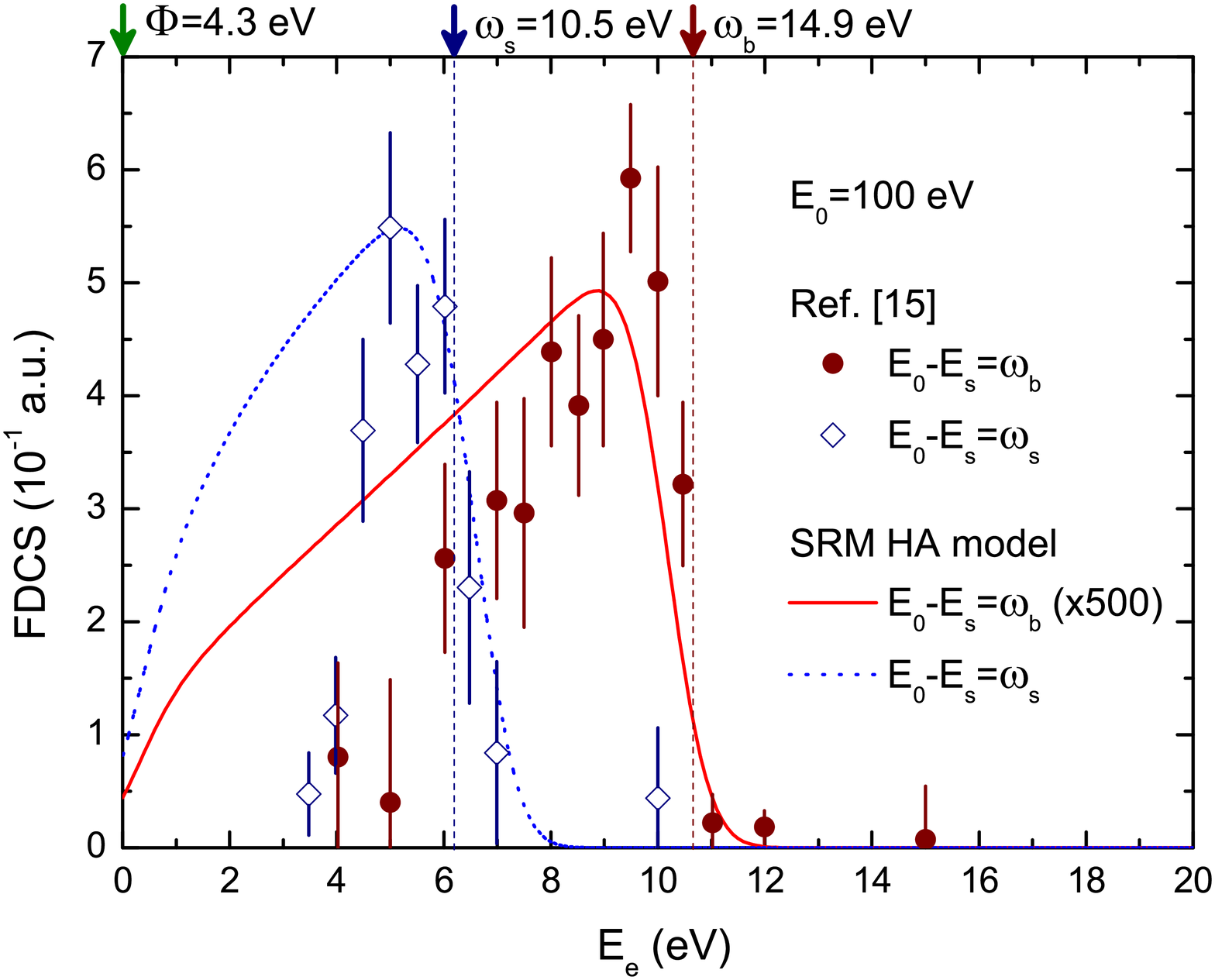}\\[1cm]
\includegraphics[scale=0.4]{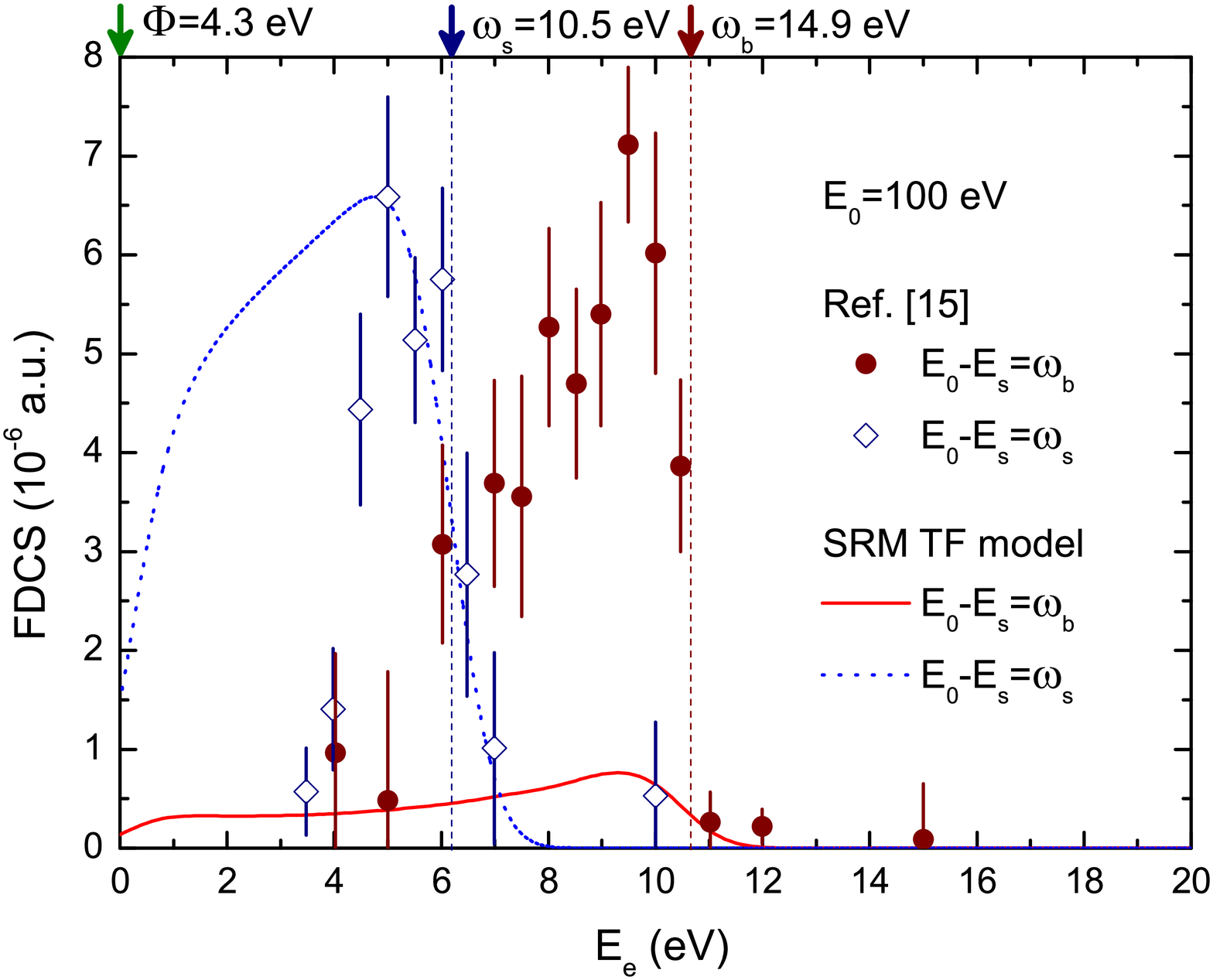}
\end{center}
\caption{\label{Al_100_SRM}(Color online) The same as in
Fig.~\ref{Al_100_Newns}, but within the SRM model.}
\end{figure}

Figs.~\ref{Al_100_Newns} and~\ref{Al_100_SRM} show the results of
the numerical calculations for the FDCS in the kinematics of
Ref.~\cite{werner08}, which were convoluted with the 2D Gaussian
function
\begin{equation}
\label{gauss08}P(E_s',E_e')=\frac{1}{2\pi\sigma_s\sigma_e}\exp\left[-\frac{(E_s'-E_s)^2}{2\sigma_s^2}-\frac{(E_e'-E_e)^2}{2\sigma_e^2}\right],
\qquad \sigma_{s(e)}=\frac{{\rm FWHM}_{s(e)}}{2\sqrt{2\ln2}},
\end{equation}
where ${\rm FWHM}_{s}=1$\,eV and ${\rm FWHM}_{e}=1.2$\,eV. As
remarked above, the absolute values were not measured in the
experiment, but the scale for the two coincidence spectra (at
$\Delta E=\omega_b$ and $\Delta E=\omega_s$) in
Ref.~\cite{werner08} is the same. In order to place the
experimental results on a common intensity scale with the theory
in each panel of Figs.~\ref{Al_100_Newns} and~\ref{Al_100_SRM},
they are normalized in such a way that the maximal experimental
and theoretical values of the FDCS corresponding to the $\Delta
E=\omega_s$ case are the same. In terms of the positions of the
peaks in the secondary-electron spectra, all calculations agree
reasonably well with experiments. However, in contrast to the
experimental results, the theoretical spectra exhibit appreciable
intensities in the region $E_e\lesssim4$\,eV. With regard to a
comparison of the intensities in the $\Delta E=\omega_b$ and
$\Delta E=\omega_s$ cases, only the RPA-IB model using the HA bulk
dielectric function provides a reasonable agreement with
experiment. Indeed, the TF approximation, both within RPA-IB and
within SRM, predicts the $\Delta E=\omega_s$ results that are by
an order of magnitude larger than the $\Delta E=\omega_b$ ones,
while the SRM model using HA yields an even much larger
discrepancy (almost three orders of magnitude). Thus, it can be
concluded that within the present theoretical approach the best
overall agreement with the experimental data of
Ref.~\cite{werner08} is found when using the RPA-IB model that
involves the HA bulk dielectric function. This comparison hints at
the suitability of the measurements to asses the reliability of
the employed dielectric response. We note, however, that further
comparisons with measurements at different geometries, as well as
on different samples, are necessary for conclusive statements on
the quality of the discussed approximations.

The experimental conditions chosen in Ref.~\cite{werner11} are
beyond the validity range and the initial scope of  our
theoretical treatment.
 In fact, in this case the theoretical FDCS vanishes for all  $\Delta E$ that were
 chosen in the experiment. The reason for this is that in the theory we do not
 account for coherent or incoherent multi-scattering events that randomize
 the momenta. As a result we are bound to
the kinematical limitation $|{\bf K}_s+{\bf K}_e-{\bf K}_0|\leq
k_F$ that stems from the  assumption of an electron ejection upon
a direct electron-electron scattering. In principle, one may lift
this limitation within the  present theoretical approach by taking
into account the effects of electron-pair diffraction
\cite{diffPRL,samarin04} and
 surface roughness. In our opinion this would be, however, insufficient
to explain the dominant, broad peak-like structures with a falling
edge at about $E_e=\omega_b-\Phi$ that are observed in the
measured coincidence spectra when $\Delta E$  exceeds largely the
$\omega_b$ value. Clearly, due to the energy balance, these
features can not be accounted for within the picture of only one
inelastic electron-electron collision involving either plasmon
decay or dynamical screening effects. The experimental
findings~\cite{werner11} thus hint at the existence of  multiple
inelastic scattering in the kinematics under study. In this
context we remark that the influence of these multiple-scattering
processes on
 the pair correlation functions was discussed and experimentally
verified for  a LiF sample in Ref.~\cite{schumann05}. In
Ref.~\cite{werner11} it was argued that multiple-scattering
processes can be viewed  as a
 Markov chain, with independent successive inelastic collisions
leading to the excitation of the surface and the bulk plasmons
followed by their decay into single electron-hole pairs (see
Ref.~\cite{werner11} for detail). We add that one cannot rule out
the scenario in which the individual inelastic events in the
Markov chain take place due to direct electron-electron scattering
resonantly enhanced at the energy transfers close to the surface-
and the bulk-plasmon frequencies.
\section{Summary and conclusions}
\label{concl}
In summary, we considered theoretically the electron energy-loss
process accompanied by the ejection of the secondary electron from
a clean metallic surface. Our study was focused on the effects of
the surface dielectric properties in the discussed process.
Restricting the analysis by the mechanism of a direct
electron-electron scattering, we took into account the
modification of the bare Coulomb potential due to the surrounding
medium by means of the inverse dielectric function. Using two
models of the inverse dielectric function of the surface, SRM and
RPA-IB, we performed numerical calculations for Al and Be with and
without accounting for the dynamical screening. We found that the
results exhibit clear plasmon features in the case of Al, while in
the case of Be these features are much less pronounced, which can
be attributed to a wider plasmon resonance in Be. For both metals,
very strong dynamical-screening effects were found.

In the present theoretical analysis we did not address the issue
of the plasmon-decay mechanism which also contributes to the
considered process. Studying this specific mechanism with the
$(e,2e)$ method can offer an opportunity for the detailed
investigation of the plasmon-decay channel into a single
electron-hole pair. For instance, the recent experimental results
from Al~\cite{werner08} showed clear surface- and bulk-plasmon
features in FDCS that were interpreted in Ref.~\cite{werner08}
within the plasmon-decay model. However, as shown by our RPA-IB HA
calculations in Fig.~\ref{Al_100_Newns}, these features can be
interpreted within the direct-scattering model as well. As stated
in the Introduction the role of the plasmon-decay mechanism is
expected to be more significant in free-electron-type metals
exhibiting a broad plasmon resonance, such as Be. However, to our
knowledge, no coincidence $(e,2e)$ measurements on such materials
have been published so far.

Thus, further experimental studies on the plasmon-assisted
$(e,2e)$ collisions at metallic surfaces are desirable in order to
shed more light on the roles of the discussed mechanisms. On the other
hand, to account for experimental observation theory must go
beyond the framework of a single inelastic collision.

\begin{acknowledgments}
We are grateful to J\"urgen Kirschner, Arthur Ernst and Sergey
Samarin for useful discussions and to Giovanni Stefani and
Alessandro Ruocco for valuable comments on their experimental
studies. This work was supported by DAAD and SFB762. K.A.K. also
acknowledges support from RFBR (Grant No.~11-01-00523-a).
\end{acknowledgments}
\begin{appendix}
\section{One-electron states}
\label{app}
Evaluation of the FDCS~(\ref{FDCS1}) requires the knowledge of the
one-electron states that correspond to the incoming (incident),
bound, and outgoing (scattered and ejected) electrons. Let us
consider a semiinfinite metallic sample filling the space in the
negative $z$ direction. Within the jellium model, the effective
one-electron potential $V$ is a steplike one, i.e.,
\begin{equation}
\label{step_pot} V=-V_0\Theta(-z).
\end{equation}
For a clean metallic material one has
\begin{equation}
V_0=\epsilon_F+\Phi,
\end{equation}
where $\Phi$ is the work function. The wave function and the
energy of the electron bound in the metal are thus given by
\begin{equation}
\label{wf_bound} \chi_{\bf k}({\bf r})=e^{i{\bf
KR}}\left\{B(k_z)e^{-\gamma
z}\Theta(z)+[e^{ik_zz}+A(k_z)e^{-ik_zz}]\Theta(-z)\right\}, \qquad
\epsilon_{\bf k}=\frac{1}{2}({\bf K}^2+k_z^2)-V_0,
\end{equation}
where
\begin{equation}
\label{wf_bound_coeff}A(k_z)=\frac{k_z-i\gamma}{k_z+i\gamma},
\qquad B(k_z)=\frac{2k_z}{k_z+i\gamma}, \qquad
\gamma=\sqrt{2V_0-k_z^2},
\end{equation}
and $\sqrt{2V_0}\geq k_z\geq0$.

Within the SRM and RPA-IB approaches the surface barrier is
impermeable for electrons in the metal. This feature can be taken
into account by calculating the wave function in
Eq.~(\ref{wf_bound}) under the assumption $V_0=\infty$, which
yields
\begin{equation}
\label{wf_IB}\chi_{\bf k}({\bf r})=2e^{i{\bf
KR}}\sin(k_zz)\Theta(-z).
\end{equation}

The incoming and the outgoing electron states with momentum ${\bf
k}=({\bf K},k_z)$ and the energy $E_{\bf k}=({\bf K}^2+k_z^2)/2$ in
the potential~(\ref{step_pot}) are given by
\begin{eqnarray}
\label{wf_in} \chi^{(+)}_{\bf k}({\bf r})&=&e^{i{\bf
KR}}\left\{[e^{ik_zz}+R(k_z)e^{-ik_zz}]\Theta(z)+D(k_z)e^{ik_z'z}e^{\alpha z}\Theta(-z)\right\} \qquad (k_z<0),\\
\label{wf_out} \chi^{(-)}_{\bf k}({\bf r})&=&e^{i{\bf
KR}}\left\{e^{ik_zz}\Theta(z)+[A_1(k_z)e^{ik_z'z}+A_2(k_z)e^{-ik_z'z}]e^{\alpha
z}\Theta(-z)\right\} \qquad (k_z>0),
\end{eqnarray}
where
\begin{eqnarray}
\label{wf_in_coeff}R(k_z)=\frac{k_z-k_z'+i\alpha}{k_z+k_z'-i\alpha},
\qquad D(k_z)=\frac{2k_z}{k_z+k_z'-i\alpha},\nonumber\\
A_1(k_z)=\frac{k_z'+k_z+i\alpha}{2k_z'},\qquad
A_2(k_z)=\frac{k_z'-k_z-i\alpha}{2k_z'}, \nonumber\\
k_z'=\mbox{sgn}(k_z)\sqrt{\frac{k_z^2+2V_0+[(k_z^2+2V_0)^2+4V_{0i}^2]^{1/2}}{2}},
\qquad \alpha=\frac{V_{0i}}{|k_z'|},\nonumber\\
\end{eqnarray}
with $V_{0i}$ being introduced as an imaginary component of the
potential,
$$
V=(-V_0+iV_{0i})\Theta(-z),
$$
to mimic the damping of the electron waves inside the metal. Using the
concept of the inelastic mean free path $\lambda$, it can be
estimated as
$$
V_{0i}=\frac{k'}{2\lambda}, \qquad k'=\sqrt{2(E_{\bf k}+V_0)}.
$$
One can calculate $\lambda$, for instance, from the
parametrization formula by Seah and Dench~\cite{seah79}, which in
the case of the elemental materials reads
\begin{equation}\label{IMFP}
\lambda=\frac{538a}{(E_{\bf k}+\Phi)^2}+0.41\sqrt{a^3(E_{\bf
k}+\Phi)},
\end{equation}
where the electron energy $E_{\bf k}+\Phi$ is measured in eV. The
average thickness of the monolayer $a$ measured in nanometers is
given by
\begin{equation}
\label{monolayer}a^3=\frac{10^{24}M}{\rho N_A},
\end{equation}
where $M$ is the atomic weight, $\rho$ is the bulk density (in
kg/m$^3$), and $N_A$ is Avogadro's number.

\end{appendix}
\pagebreak
%


\begin{thebibliography}{99}
%
\bibitem{eels_book} R.~Byrdson, \emph{Electron Energy Loss
Spectrosocpy} (BIOS Scientific, Oxford, 2001).
%
\bibitem{samarin95}J. Kirschner, O. M. Artamonov, and S. N. Samarin, Phys.
Rev. Lett. {\bf 75}, 2424 (1995).
%
\bibitem{schumann05}F. O. Schumann, J. Kirschner, and J. Berakdar, Phys. Rev.
Lett. {\bf 95}, 117601 (2005).
%
\bibitem{schumann10}F. O. Schumann, C. Winkler, J. Kirschner, F. Giebels, H. Gollisch, and R.
Feder, Phys. Rev. Lett. {\bf 104}, 087602 (2010).
%
%
\bibitem{weigold_book}E. Weigold and I. E. McCarthy, \emph{Electron Momentum Spectroscopy} (Kluwer, New York, 1999).
%
%
\bibitem{prl99} J. Berakdar, Phys. Rev. Lett. {\bf 83}, 5150 (1999).
%

\bibitem{prl2000} S. N. Samarin, J. Berakdar, O. Artamonov, and J. Kirschner, Phys.
Rev. Lett. {\bf 85}, 1746 (2000).
%
\bibitem{Halle_PRB2002}A. Morozov, J. Berakdar, S. N. Samarin, F. U. Hillebrecht, and J. Kirschner, Phys. Rev. B {\bf 65}, 104425 (2002).

%
%
\bibitem{das97}J. Berakdar, M. P. Das, Phys. Rev. A {\bf 56}, 1403 (1997).
%
%
\bibitem{diffPRL}J. Berakdar, S. N. Samarin, R. Herrmann, and J. Kirschner, Phys. Rev. Lett. {\bf 81}, 3535 (1998).
%
%
\bibitem{gollisch2001} H. Gollisch, T. Scheunemann, and R. Feder, Solid State Commun. {\bf 117}, 691 (2001).
%
%
\bibitem{berakdarSOLSTCOM} J. Berakdar, H. Gollisch, and R. Feder, Solid State Commun. {\bf 112}, 10587 (1999).
%
\bibitem{serg2006} S. Samarin, O. M. Artamonov, A. D. Sergeant, R. Stamps, and J. F. Williams, Phys. Rev. Lett. {\bf 97}, 096402 (2006).
%
\bibitem{serg2011} S. Samarin, O. M. Artamonov, V. N. Petrov, M. Kostylev, L. Pravica, A. Baraban, and J. F. Williams, Phys. Rev. B {\bf 84}, 184433 (2011).
%
%
\bibitem{werner08} W. S. M. Werner, A. Ruocco, F. Offi, S. Iacobucci, W. Smekal, H. Winter,
and G. Stefani, Phys. Rev. B {\bf 78}, 233403 (2008).
%
\bibitem{werner11} W. S. M. Werner, F. Salvat-Pujol, W. Smekal, R. Khalid, F. Aumayr, H. St\"ori, A. Ruocco, and G.
Stefani, Appl. Phys. Lett. {\bf 99}, 184102 (2011).
%
%
\bibitem{kouzakov03}K. A. Kouzakov and J. Berakdar, Phys. Rev. A {\bf 68}, 022902
(2003).
%
%
\bibitem{arpes} \emph{Angle-Resolved Photoemission: Theory and
Applications}, edited by S. D. Kevan (Elsevier, Amsterdam, 1992).
%
\bibitem{pes}S.~H\"ufner, \emph{Photoelectron Specrosocopy - Principles and
Applications, 3rd ed.} (Springer, Berlin, 2003).
%
\bibitem{fano63} U. Fano, Annu. Rev. Nucl. Sci. {\bf 13}, 1 (1963).
%
\bibitem{schockley33}W. Shockley, Phys. Rev. {\bf 56}, 317 (1939).
%
\bibitem{tamm32}I. Tamm, Phys. Z. Sov. Union {\bf 1}, 733 (1932).
%
\bibitem{samarin04} S. Samarin, J. Berakdar, A. Suvorova, O. M. Artamonov, D. K. Waterhouse, J. Kirschner, and J. F.
Williams, Surf. Sci. {\bf 548}, 187 (2004).
%
%
\bibitem{ritchie57}R. H. Ritchie, Phys. Rev. {\bf 106}, 874 (1957).
%
\bibitem{silkin04}V. M. Silkin, A. Garc\'ia-Lekue, J. M. Pitarke, E. V. Chulkov, E.
Zaremba, and P. M. Echenique, Europhys. Lett. {\bf 66}, 260
(2004).
%
\bibitem{pitarke04}J. M. Pitarke, V. U. Nazarov, V. M. Silkin, E. V. Chulkov, E.
Zaremba, and P. M. Echenique, Phys. Rev. B {\bf 70}, 205403
(2004).
%
\bibitem{nature07} B. Diaconescu, K. Pohl, L. Vattuone, L. Savio, P. Hofmann, V. M. Silkin,
J. M. Pitarke, E. V. Chulkov, P. M. Echenique, D. Far\'ias, and M.
Rocca, Nature (London) {\bf 448}, 57 (2007).
%
\bibitem{chung77}M. S. Chung and T. E. Everhart, Phys. Rev. B {\bf
15}, 4699 (1977).
%
\bibitem{pines53}D. Bohm and D. Pines, Phys. Rev. {\bf 92}, 609 (1953)

\bibitem{srm66}R. H. Ritchie and A. L. Marusak, Surf. Sci. {\bf
4}, 234 (1966).
%
\bibitem{newns70}D. M. Newns, Phys. Rev. B {\bf 1}, 3304 (1970).
%
\bibitem{kouzakov02}K. A. Kouzakov and J. Berakdar, Phys. Rev. B {\bf 66}, 235114
(2002).
%
\bibitem{leed} J. B. Pendry, \emph{Low Energy Electron Diffraction} (Academic, New York, 1974).
%
\bibitem{luders01}M. L\"uders, A. Ernst, W. M. Temmerman, Z. Szotek, and P. J.
Durham, J. Phys.: Condens. Matter {\bf 13}, 8587 (2001).
%
\bibitem{echenique92}F. J. Garc\'ia de Abajo and P. M. Echenique,
Phys. Rev. B {\bf 46}, 2663 (1992).
%
\bibitem{lindhard}J. Lindhard, K. Dan. Vidensk. Selsk. Mat. Fys. Medd. {\bf 28}, No.8
(1954).
%
\bibitem{bechstedt83}F. Bechstedt, R. Enderlein, and D. Reichardt, Phys. Status
Solidi B {\bf 117}, 261 (1983).
%
\bibitem{horring85}N. J. Morgenstern Horing, E. Kamen, and H.-L. Cui, Phys.
Rev. B {\bf 32}, 2184 (1985).
%
\bibitem{ashcroft_book}N. W. Ashcroft and N. D. Mermin, {\it Solid State Physics} (Hault-Saunders, Tokyo, 1981).
%
\bibitem{bloch33}F. Bloch, Z. Phys. {\bf 81}, 363 (1933); Helv.
Phys. Acta {\bf 7}, 385 (1934).
%
\bibitem{halevi95}P. Halevi, Phys. Rev. B {\bf 51}, 7497 (1995).
%
%
\bibitem{seah79}M. P. Seah and W. A. Dench, Surf. Interface Anal.
{\bf 1}, 2 (1979); M. P. Seah, \emph{ibid.} {\bf 9}, 85 (1986).
%
%
\end{thebibliography}
\end{document}